\documentclass[aps,amsmath,prd,nofootinbib,preprint,preprintnumbers,eqsecnum]{revtex4}

\usepackage[normalem]{ulem}
\usepackage{amsmath}
\usepackage{enumerate}
\usepackage{amsfonts}
\usepackage{epsfig}
\usepackage{yfonts}
\usepackage{subfigure}

\usepackage{bbm}
\usepackage{epstopdf}

\usepackage{mathtools}

\usepackage[
	  pagebackref=false,
	  colorlinks=true,
      linkcolor=blue,
      urlcolor=blue,
      filecolor=black,
      citecolor=red,
      pdfstartview=FitV,
      pdftitle={},
        pdfauthor={},
        pdfsubject={},
        pdfkeywords={},
        pdfpagemode=None,
        bookmarksopen=true
      ]{hyperref}

\usepackage{graphicx}
\usepackage{bm}

\def\({\left(}
\def\){\right)}
\def\[{\left[}
\def\]{\right]}
\def\<{\langle}
\def\>{\rangle}

\newcommand\half{{\ensuremath{\frac{1}{2}}}}
\newcommand\p{\ensuremath{\partial}}

\newcommand{\be}{\begin{equation}}
\newcommand{\ee}{\end{equation}}
\newcommand{\bea}{\begin{eqnarray}}
\newcommand{\eea}{\end{eqnarray}}
\newcommand{\bwt}{\begin{widetext}}
\newcommand{\ewt}{\end{widetext}}

\newcommand{\bi}{\begin{itemize}}
\newcommand{\ei}{\end{itemize}}
\newcommand{\ben}{\begin{enumerate}}
\newcommand{\een}{\end{enumerate}}
\newcommand{\bca}{\begin{cases}}
\newcommand{\eca}{\end{cases}}
\newcommand{\bln}{\begin{align}}
\newcommand{\eln}{\end{align}}
\newcommand{\bst}{\begin{split}}
\newcommand{\est}{\end{split}}

\newcommand\al{{\alpha}}
\newcommand\ep{\epsilon}
\newcommand\sig{\sigma}
\newcommand\Sig{\Sigma}
\newcommand\lam{\lambda}
\newcommand\Lam{\Lambda}
\newcommand\om{\omega}
\newcommand\Om{\Omega}

\newcommand\ga{{\ensuremath{{\gamma}}}}
\newcommand\Ga{{\ensuremath{{\Gamma}}}}
\newcommand\de{{\ensuremath{{\delta}}}}
\newcommand\De{{\ensuremath{{\Delta}}}}

\newcommand\ov{\over}
\newcommand\ha{{\half}}

\def\le{\left}
\def\ri{\right}

\newcommand\sA{{\ensuremath{{\mathcal A}}}}

\newcommand\sC{{\ensuremath{{\mathcal C}}}}

\newcommand\sH{{\ensuremath{{\mathcal H}}}}
\newcommand\sL{{\ensuremath{{\mathcal L}}}}

\newcommand\sR{{\ensuremath{{\mathcal R}}}}

\newcommand\sS{{\mathcal S}}

\renewcommand{\Im}{\textrm{Im}\,}

\newcommand{\open}{\left( \begin{array}{cc} }
\newcommand{\opent}{\left( \begin{array}{ccc} }
\newcommand{\opens}{\left( \begin{array}{cccc} }
\newcommand{\close}{\end{array}\right)}

\newcommand{\ka}{{\kappa}}

\newcommand{\sds}{s^{(\Sig)}_d}

\begin{document}

\title{Probing renormalization group flows using entanglement entropy}

\preprint{MIT-CTP 4500}

\author{ Hong Liu}
\affiliation{Center for Theoretical Physics, \\
Massachusetts
Institute of Technology,
Cambridge, MA 02139 }
\author{M\'ark Mezei}
\affiliation{Center for Theoretical Physics, \\ Massachusetts Institute of Technology,
Cambridge, MA 02139 }

\begin{abstract}

\noindent In this paper we continue the study of renormalized entanglement entropy introduced in~\cite{Liu:2012eea}. In particular, we investigate its behavior near an IR fixed point using holographic duality. We develop techniques which,  for any static holographic geometry, enable us to extract the large radius expansion of the entanglement entropy  for a spherical region.
 We show that for both a sphere and a strip, 
the  approach of the renormalized entanglement entropy to the IR fixed point value contains 
a contribution that depends on the whole RG trajectory. Such a contribution is dominant, when the leading 
irrelevant operator is sufficiently irrelevant. For a spherical region such terms can be anticipated 
from a geometric expansion, while for a strip whether these terms have geometric origins remains to be seen.

\end{abstract}


\maketitle

\tableofcontents

\section{Introduction and summary}


In renormalizable field theories, the entanglement entropy (EE) for a spatial region is divergent in the continuum limit, with the leading divergence given by the so-called area law \cite{Bombelli:1986rw,Srednicki:1993im}:
\be \label{areal}
S^{(\Sig)} =\#\, {\sA_\Sig\ov \delta^{d-1}}+\cdots \ ,
\ee
where $\delta$ is a short-distance cutoff, $d$ is the number of spacetime dimensions, $\sA_\Sig$ is the area of the entangling surface $\Sig$, and the dots stand for less singular terms. Equation~\eqref{areal} can be interpreted as coming from degrees of freedom at the cutoff scale $\de$ near $\Sig$. 

More generally, for a smooth $\Sig$,  one expects that  local contributions (including all divergences) 
near $\Sig$ to the entanglement entropy  can be 
written in terms of local geometric invariants of $\Sig$~\cite{tarun,Liu:2012eea}
\be \label{locE}
S_{\rm local}^{(\Sig)} = \int_\Sig {d^{d-2} \sig} \,  \sqrt{h} F (K_{ab}, h_{ab}) \
\ee
where $\sig$ denotes coordinates on $\Sig$, $F$ is a sum of all possible local geometric invariants formed from the induced metric $h_{ab}$ and extrinsic curvature $K_{ab}$ of $\Sig$. For a {\it scalable} surface $\Sig$ of size $R$,\footnote{A scalable surface can be specified by a size $R$ and a number of dimensionless 
parameters characterizing the shape.}  the local contribution~\eqref{locE} should then have the following geometric expansion 
\be \label{divv}
S_{\rm local}^{(\Sig)} = a_1 R^{d-2} + a_2 R^{d-4} + \cdots 
\ee
with the first term coinciding with~\eqref{areal}. 
In~\eqref{divv} terms with positive exponents of $R$ are expected to be divergent,\footnote{We assume a continuous regularization in which  the size $R$ can unambiguously defined.} while
$a_n$ is finite, when the corresponding exponent of $R$ is negative.  

The area law~\eqref{areal} and other subleading divergences in~\eqref{divv} indicate that EE is dominated by physics at the cutoff scale and thus is not a well defined observable in the continuum limit.  This UV-sensitivity makes it difficult to extract long range correlations from EE. A standard practice is 
to subtract the divergent part by hand. This may not be sufficient to remove all the short-distance dependence, and is often ambiguous.  For example,  consider the entanglement entropy of a disk of radius $R$ in the vacuum of a (2+1)-dimensional free massive scalar field theory. It was obtained in~\cite{Hertzberg:2010uv,Huerta:2011qi} that for $m R \gg 1$, the entanglement entropy has the behavior 
\be \label{frema}
S (R)= \# {R \ov \de} - {\pi \ov 6} m R - {\pi \ov 240}{1 \ov mR} + \cdots \ .
\ee
Subtracting the divergent $\# {R \ov \de}$ piece by hand, from the second term in~\eqref{frema}  one finds that in the IR limit ($R \to \infty$), the resulting expression approaches $-\infty$. This result appears to be in conflict with the expectation that in the IR limit the system should have no correlations. 
Ideally, we would have liked EE to go to zero. To understand what is going on, note that the second term in~\eqref{frema} also has the form of an area law  ${R / \tilde \de}$ with  $\tilde \de \sim {1 \ov m}$, and  thus can be interpreted as coming from physics at scale ${1 \ov m}$, which is still short-scale physics compared to the IR scale $R \to \infty$. 
A related observation is that the second term in~\eqref{frema} is in fact ambiguous in the continuum limit, as its coefficient can be modified by the following redefinition of $\de$ 
\be 
\de \to \de \le(1 + c\, m \de + \cdots \ri) \ 
\ee
with $c$ some constant. 

In~\cite{Liu:2012eea} we introduced the renormalized entanglement entropy (REE) 
 \be \label{Scen}
 \sS_d^{(\Sig)} (R) = \bca
  {1 \ov  (d-2)!!}  \le(R {d \ov dR} - 1 \ri) \le(R {d \ov dR} -3 \ri)  \cdots \le(R {d \ov dR} -(d-2) \ri) S^{(\Sig)}(R) & {\rm d \; odd} \cr
  {1 \ov  (d-2)!!}  R {d \ov dR}  \le(R {d \ov dR} -2 \ri)  \cdots \le(R {d \ov dR} -(d-2) \ri) S^{(\Sig)}(R) & {\rm d \; even}
      \eca   \  
    \ee
which was designed to remove all divergent terms in~\eqref{divv}.   
It was shown there that  the REE has the following desired properties:\footnote{The differential operator~\eqref{Scen} can be applied to the R\'enyi entropies and the following statements also apply to renormalized R\'enyi entropies.}  
\ben 

\item  It is unambiguously defined  in the continuum limit.  

\item For a CFT it is given by a $R$-independent constant  $\sds$. \label{im:2}

\item For a renormalizable quantum field theory, it 
interpolates between the values $s^{(\rm \Sig,UV)}_d$ and $s^{(\rm \Sig,IR)}_d$ of the UV and IR fixed points as $R$ is increased from zero to infinity.  
\label{im:3}

\item It is most sensitive to degrees of freedom at scale $R$. \label{im:4}

\een 
For example, applying~\eqref{Scen} to~\eqref{frema} we find that the differential operator in~\eqref{Scen} (for $d=3$) removes the first two terms in~\eqref{frema} and changes the sign of the last term, resulting
\be 
\sS_3 (R) = + {\pi \ov 120} {1 \ov mR} + \cdots , \qquad m R \to \infty
\ee
which monotonically {\it decreases} to zero at large distances as desired. 

For a general quantum field theory the REE can  be interpreted as characterizing 
entanglement at scale $R$.  In particular, the $R$-dependence can be interpreted as describing the renormalization group (RG) flow of entanglement entropy with distance scale. In~\cite{Liu:2012eea}, it was conjectured that in three spacetime dimension the REE for a sphere 
 $\sS_{3}^{\rm sphere}$ is monotonically decreasing and non-negative for the vacuum of Lorentz invariant, unitary QFTs,  providing a central function for the F-theorem conjectured previously in~\cite{Myers:2010xs,Jafferis:2011zi}. The monotonic nature of  $\sS_{3}^{\rm sphere}$, 
and thus the F-theorem, was subsequently proved in~\cite{Casini:2012ei}. 
In $(1+1)$-dimension, $\sS_2$ reduces to an expression previously considered in~\cite{Casini:2004bw}, where 
its monotonicity was also established. There are, however, some indications~\cite{Liu:2012eea} that in four spacetime dimensions $\sS_{4}^{\rm sphere}$ is neither monotonic nor non-negative.

More generally, regardless of whether it is monotonic, REE  provides a new set of observables to probe RG flows.\footnote{See~\cite{Latorre:2004pk,Riera:2006vj} for other ideas for probing RG flows using entanglement entropy.} From REE, one can introduce an ``entropic function'' 
defined in the space of couplings (or in other words the space of theories)
\be \label{cfun}
\sC^{(\Sig)} (g^a (\Lam)) \equiv \sS^{(\Sig)} \le(R \Lam, g^a (\Lam) \ri) \bigr|_{R = {1 \ov \Lam}}
= \sS^{(\Sig)} \le(1, g^a (\Lam) \ri)
\ee
where $g^a (\Lam)$ denotes collectively all couplings
and $\Lam$ is the RG energy scale. Given that $\sS^{(\Sig)}$ is a measurable quantity,  it should satisfy the Callan-Symanzik equation
\be
 \Lam {d \sS^{(\Sig)} \le(R \Lam, g^a (\Lam) \ri) \ov d\Lam} = 0 \ ,
\ee
which leads to
\be
\Lam {d \sC^{(\Sig)} (g^a (\Lam)) \ov d \Lam} 
= - R{d \sS^{(\Sig)} \le(R \Lam, g^a (\Lam) \ri) \ov d R} \biggr|_{R = {1 \ov \Lam}} \ .
\ee
The $R$-dependence of $\sS^{(\Sig)}$ is translated into the running of $\sC^{(\Sig)} (g (\Lam))$ in the space of couplings, with $R \to 0$ and $R \to \infty$ limits correspond to approaching UV and IR fixed points of RG flows. 
At a fixed point $g_*$, $\sC^{(\Sig)}  (g_*) = s_d^{(\Sig)} $ and the monotonicity of $\sS^{(\Sig)}$ with respect to $R$ translates to the monotonicity of $\sC^{(\Sig)}$ with respect to $\Lam$. 
 
For $\Sig$ being a sphere, some partial results were obtained earlier in~\cite{Liu:2012eea,Klebanov:2012yf} for the small and large $R$ behavior of REE (or equivalently for $\sC$ near a UV and IR fixed point) in holographic theories. 
From now on we will focus on a spherical region and suppress the superscript ${(\Sig)}$ on $\sS$ and $\sC$. 
 For a (UV) fixed point perturbed by a relevant operator of dimension $\De < d$,  it was found that 
\be \label{smlRB}
\sS_d  (R) = s_d^{(\rm UV)} - 
A (\De)  (\mu R)^{2 (d- \De)} + \cdots ,\quad R \to 0
\ee
where $\mu$ is a mass scale with the relevant (dimensional) coupling given by
$g = \mu^{d-\De}$, and $A(\De)$ is some {\it positive} constant. The above equation leads to an entropic function 
given by
\be \label{smarb1}
\sC_d (g) =  s_d^{( \rm UV)} - A (\De) g^2_{eff} (\Lam), \qquad \Lam \to \infty
\ee
where $g_{eff} (\Lam) =g  \Lam^{\De -d}$ is the effective dimensionless coupling at scale $\Lam$. 
Equation~\eqref{smarb1} has a simple interpretation that the leading UV behavior of the entropic function is controlled by the two-point correlation function of the corresponding relevant operator.  We expect this result to be valid also outside holographic systems. This appears to be also consistent with general arguments from conformal perturbation theory~\cite{Cardy:2010zs}. 
It is curious, however, that low dimensional free theories defy this expectation. For example in 
$d=2$, 
as $R \to 0$~\cite{Casini:2005rm}
\bea \label{fee1}
&& {\rm free \; scalar:} \qquad 
\sS_2 (R)  
=\frac13 + {1\ov \log \le(m^2 R^2\ri)}+\cdots  
 \\
 \label{fee2}
&& {\rm Dirac \; fermion}: \qquad 
\sS_2 (R)  
=\frac13 - 4 m^2 R^2\, \log^2 \le(m^2 R^2\ri)+\cdots \ ,
\eea
while for a $d=3$ free massive scalar~\cite{Klebanov:2012va} ruled out the $m^4 R^4$ short distance behavior based on numerics.\footnote{Note that the relevant deformation of the massless scalar UV fixed point, $\phi^2$ has dimension $\Delta=1$, hence~\eqref{smlRB} would predict an $m^4 R^4$ behavior.} 

Near an IR fixed point, it was argued in~\cite{Liu:2012eea} that the large $R$ behavior of $\sS (R)$ should have the form 
\bea 
\sS_d (R) &=&  s_d^{(\rm IR)}  +
{B (\tilde \De) \ov (\tilde \mu R)^{2 (\tilde \De -d)} }+ \cdots \cr 
& & + \bca 
{s_1 \ov \tilde \mu R} + {s_3 \ov (\tilde \mu R)^3} + \cdots & {\rm odd  } \; d \cr
{s_2 \ov (\tilde \mu R)^2} + {s_4 \ov (\tilde \mu R)^4} + \cdots & {\rm even } \; d 
\eca ,\quad R \to \infty \ ,
 \label{irex}
  \eea
where $\tilde \De > d$ is the dimension of the leading irrelevant operator, $\tilde \mu$ is a mass scale 
characterizing the irrelevant perturbation, and $B (\tilde \De)$ is a constant.  The first line, similar to~\eqref{smlRB}, has a natural interpretation in terms of conformal perturbations of the 
IR fixed point.  The coefficient $B (\tilde \De)$ is expected to depend only on physics of the IR fixed point. In terms of irrelevant coupling $\tilde g = \tilde \mu^{d- \De}$ corresponding to the leading irrelevant operator,  
equation~\eqref{irex} leads to
\bea 
\sC (\Lam) &=&  s_d^{(\rm IR)}  +
B (\tilde \De) \tilde g_{eff}^2 (\Lam)+ \cdots \cr 
& & + \bca 
s_1 \tilde g^{1 \ov \tilde \De -d}_{eff} (\Lam) + \cdots & {\rm odd  } \; d \cr
s_2 \tilde g^{2 \ov \tilde \De -d}_{eff} (\Lam) + \cdots & {\rm even } \; d 
\eca , \quad \Lam \to \infty  \ ,
 \label{irex1}
  \eea
where $\tilde  g_{eff} (\Lam) = \tilde g \Lam^{\tilde \De - d}$ is the effective dimensionless coupling at scale 
$\Lam$. It is amusing that the ``analytic''  contributions in $1/R$ in~\eqref{irex} lead to non-analytic dependence on the coupling while non-analytic contributions in $1/R$ lead to analytic dependence on the coupling. Note the first line dominates for 
\be \label{NotTooIrrellevant}
\tilde \De <  \bca  d + \ha & {\rm odd} \; d \cr
           d + 1 & {\rm even} \; d
           \eca
           \ee
i.e. if the leading irrelevant operator is not too irrelevant. Note in this range $B(\tilde\De)>0$. The second line of~\eqref{irex}-\eqref{irex1}
can be expected from~\eqref{divv}: the contributions of any degrees of freedom at some lengths scale $\ell \ll R$
should have an expansion of the form~\eqref{divv}. Thus the coefficients $s_n$ are expected to depend on the RG trajectory from the cutoff scale $\de$ to $R$.\footnote{Since here we consider the $R \to \infty$ limit 
$s_n $ should thus depend on the full RG trajectory from $\de$ to $\infty$.}

Support for~\eqref{irex} was provided in~\cite{Liu:2012eea} by examining holographic RG flows between two closely separated fixed points. In this paper we prove~\eqref{irex} for all Lorentz invariant holographic flows
with an IR conformal fixed point, which is described on the gravity side by a domain wall geometry interpolating 
between two AdS spacetimes of different cosmological constant.  
In particular, we show that $B(\tilde \De)$ is the same as that obtained earlier for RG flows between two closely separated fixed points; this is consistent with the expectation that it should only depend on the physics at the IR fixed point.  We obtain a general expression for $s_1$ in $d=3$  in terms of an integral of the spacetime metric over the full spacetime. With more diligence, other coefficients in generic $d$ dimensions can be straightforwardly obtained using our techniques, although we will not determine them here. 

In addition to domain wall geometries, we also consider a class of geometries, which are singular
in the IR. These correspond to either gapped systems, or 
systems whose IR fixed point does not have a gravity description (or has degrees of freedom smaller than 
$O(N^2)$). We will see that for these geometries the asymptotic behavior of REE provides a simple diagnostic 
of IR gapless degrees of freedom. 

While in this paper we focus on the vacuum flows, the techniques we develop can be used to obtain the large $R$ expansion of the entanglement entropy for generic static holographic geometries, including  nonzero temperature and  chemical potential. As an illustration we study the behavior of extremal surfaces in a general black hole geometry in the large size limit. We also show that, in this limit, for {\it any shape} of the entangling surface the leading behavior of the EE is the thermal entropy. While this result is anticipated, a general holographic proof appears to be lacking so far.



For $d=2,3$, the monotonicity of $\sS_d$ in $R$ leads to a monotonic $\sC_d$ in coupling space, i.e. $\sC_d$ is a c-function. Equations~\eqref{fee1}--\eqref{fee2} show that for a free massive field, $\sC_2$ is not stationary near the UV fixed point, and neither is $\sC_3$ for a free massive scalar field, as pointed out in~\cite{Klebanov:2012va}. From~\eqref{irex1} we see that $\sC_d$ is in fact generically non-stationary near an IR fixed point
for $\tilde \De -d > \ha$ ($\tilde \De -d > 1$) for odd (even) dimensions. 
The physical reason behind the non-stationarity is simple: while the contribution from degrees of freedom at short length scales 
are suppressed in $\sS_d$, they are only suppressed as a fixed inverse power of $R$, and are the dominant subleading contribution, when the leading irrelevant operator is sufficiently irrelevant. 
The non-stationarity of $\sS$ (or $\sC$) is independent of the monotonic nature of $\sS$~(or $\sC$) and 
should not affect the validity of c- or F-theorems. 
In contrast to the Zamolodchikov c-function~\cite{Zamolodchikov:1986gt}, which is stationary, in our opinion, 
the  non-stationarity of $\sC$ should be considered as an advantage, as it provides a more sensitive 
probe of RG flows. For example, from~\eqref{irex1} by merely examining the leading approach to an IR 
fixed point, one could put constraints on the dimension of the leading irrelevant operator. 

While in this paper we will be mainly interested in taking the entangling surface to be a sphere of radius $R$, for comparison we also examine the IR behavior for a strip. Since the boundary of a strip is not scalable, the definition~\eqref{Scen} has to be modified. Consider a strip 
\be \label{defstrip}
x_1 \in (-R, R), \quad x_i \in (0, \ell), \quad i =2, \cdots, d-1
\ee
where for convenience we have put other spatial directions to have a finite size $\ell \to \infty$. 
Note that due to translational symmetries of the entangled region in $x_i$ directions, the EE should have an extensive dependence on $\ell$, i.e. it should be proportional to $\ell^{d-2}$. 
Furthermore, for the boundary of a strip the extrinsic curvature and all tangential derivatives vanish. Hence we conclude that the only divergence is the area term
\be
S_{\rm strip} (R)=\ell^{d-2}  \le({\# \ov \delta^{d-2}}+{\rm finite} \ri) \ .
\ee
In particular, the divergent term should be $R$-independent. 
This thus motivates us to consider $R {dS \ov d R}$, which should be finite and devoid of any cutoff dependent ambiguities. Given that all the dependence in $S$ on $\ell$ comes from the over factor $\ell^{d-2}$, it is convenient 
to introduce dimensionless quantity $\sR_d$  defined by 
\be \label{scke}
 R {dS \ov dR} \equiv {\ell^{d-2} \ov R^{d-2}} \sR_d (R) \ .
\ee
This quantity was considered earlier in~\cite{Ryu:2006ef,Myers:2012ed}. For a CFT there is no scale other than $R$, hence $\sR_d$ should be a $R$-independent constant, which can be readily extracted from expressions in~\cite{Ryu:2006bv,Ryu:2006ef}. For a general QFT, $\sR_d$ should be a dimensionless combination of $R$ and other possible mass scales of the system. 

Calculating $\sR_d$ for a domain wall geometry describing flows among two conformal fixed points, we find an interesting surprise. The second line of~\eqref{irex} can be understood from a local curvature expansion 
associated with a spherical entangling surface. Such curvature invariants  altogether vanish for a strip 
and thus one may expect that for a strip only the first line of~\eqref{irex} should be present. We find instead find that
$\sR_d$ has the large $R$ behavior 
\bea  \label{stipREE}
\sR_d (R) &=& \sR^{(IR)}_d +
     c (\tilde \De)  (\tilde \mu R)^{-2 (\tilde \De -d)} +\dots  \cr
     &&+t_d (\tilde \mu R)^{-d} +\dots \ , 
\eea
where $\sR^{(IR)}_d$ is $R$-independent constant characterizing the IR fixed point, $c (\tilde \De)$ is a constant which depends only on the IR data, while the constant $t_d$ involves an integral over the whole 
radial direction, signaling that this term receives contributions from degrees of freedom of all length scales. Note that similarly to the sphere case, the terms in the second line is the leading approach to the IR fixed point value for  $\tilde \De >  {3d / 2} $.
Note that the terms we find come from the following terms in $S_{\rm strip} (R)$:
\bea
S_{\rm strip} (R)&=&\ell^{d-2}  \le({\# \ov \delta^{d-2}}-{\sR^{(IR)}_d\ov (d-2)\,R^{d-2}}-{c (\tilde \De)\ov (2 \tilde \De -d-2)\tilde \mu ^{2 (\tilde \De -d)}}\,  R^{-(2 \tilde \De -d-2)}+\dots\ri. \cr
&& \le.-{t_d\ov 2(d-1) \tilde \mu^d} (\tilde \mu R)^{-2(d-1)}+\dots\ri) \ .
\eea
It would be interesting to see, whether it is possible to identify a geometric origin for the terms in the second line.

The paper is organized as follows.  In Sec.~\ref{sec:setup} we discuss 
the holographic geometries to be considered, and outline a general strategy to 
obtain the large $R$ expansion of REE for a spherical region for 
 generic holographic geometries. In Sec.~\ref{sec:gapped} we consider holographic theories which are gapped or whose IR fixed point does not have a good gravity description.  In Sec.~\ref{sec:scal} we elaborate more
 on the physical interpretation of such geometries and consider some explicit examples. In Sec.~\ref{sec:domain} we consider domain wall geometries with an IR conformal fixed point. We conclude in Sec.~\ref{sec:bh} with some applications of the formalism to the black hole geometry.

\section{Setup of the calculation and general strategy} \label{sec:setup}

In this section we describe the basic setup for our calculations and outline the general strategy.

\subsection{The metric}

The RG flow of a Lorentz-invariant holographic system in the vacuum can be described by a metric of the form 
\be \label{1}
ds^2  =  {L^2 \ov z^2} \le(- dt^2 + d \vec x^2 + {dz^2 \ov f(z)} \ri) \ ,
\ee
where $L$ is the AdS radius and  near the boundary
\be 
f (z) \to 1,  \qquad z \to 0 \ . 
\ee
The null energy condition requires $f$ to be monotonically increasing. 
The IR behavior, as $z \to \infty$ can then have the following  two possibilities:  

\ben 

\item  $f$ approaches a finite constant
\be \label{pep1}
f(z)  \to  {L^2 \ov L_{IR}^2}\equiv  f_\infty  > 1 , \qquad z \to \infty      \ .
\ee
In this case, the IR geometry is given by AdS with radius $L_{IR} < L $, and thus the system
flows to an IR conformal fixed point. 
 Near the IR fixed point, i.e. $z \to \infty$, $f$ can be expanded as 
 \be \label{find1}
 f (z) =f_\infty \le(1 - {1 \ov (\tilde \mu z)^{2 \tilde \al}} + \cdots \ri) \ ,
 \ee
where $\tilde \al = \tilde \De - d$, with $\tilde\De$ being the dimension of the leading irrelevant perturbing operator at the IR fixed point, and $\tilde \mu$ is a mass scale characterizing irrelevant perturbations.

\item The spacetime becomes singular at $z=\infty$:
\be \label{singf}
f(z) = a z^{n}+\cdots, \qquad a> 0, \quad n > 0 \ .
\ee 
Due to the singularity at $z=\infty$, one might be concerned, whether one could trust 
the holographic entanglement entropy obtained in such a geometry. We will see, however, that the results obtained 
in this paper only depend on the existence of the scaling behavior~\eqref{singf} for a certain range of $z$ and are insensitive to how the singularity at $z = \infty$ is resolved. 

Since $n > 0$, the singularity lies at a finite proper distance away and the naive expectation is the corresponding IR phase should be gapped. 
As we will discuss later, it turns out this is only true for $n > 2$, an example of which 
is  the GPPZ flow~\cite{Girardello:1999hj}. 
For $n < 2$, the story is more intricate and there exist gapless modes 
in the IR. Below we will refer to $n <  2$ geometries scaling geometries, examples of which include 
the near horizon geometries of D1, D2 and D4-branes. In these examples, the IR fixed point either does not have a 
good gravity description (like in the case of D1 or D4 branes) or the number of degrees of freedom at the IR fixed point scales with $N$ with a lower power than $N^2$ (like in the case of D2 branes, where the IR description is in terms of M2 branes giving $N^{3/2}$ degrees of freedom). Thus one should interpret the scaling region~\eqref{singf} 
as describing an intermediate scaling regime of the boundary theory before the true IR phase is reached.

\een
In our subsequent discussion we will assume that there exists a crossover scale $z_{CO}$ 
such that~\eqref{find1} or~\eqref{singf} is valid for 
\be 
z \gg z_{CO} \ .
\ee

While in this paper we will be focusing on vacuum solutions (i.e. with Lorentz symmetry), since the holographic computation of the entanglement entropy for a static system only depends on the spatial part of the metric~\cite{Ryu:2006bv}, the techniques we develop in this paper for calculating the large $R$ behavior of the REE also apply to a more general class of 
metrics of the form 
\be \label{2}
ds^2  =  {L^2 \ov z^2} \le(- g(z) dt^2 + d \vec x^2 + {dz^2 \ov f(z)} \ri)  \ .
\ee
This is in fact the most general metric describing a translational and rotational invariant boundary system including all finite temperature and finite chemical potential solutions.
$g$ does not directly enters the computation of the REE. Its presence is felt in the more general behavior allowed for $f$;
 the null energy condition  no longer requires $f$ to be monotonically increasing. For example, for a black hole solution $f$ decreases from the boundary value $1$ to 
 zero at the horizon. The null energy condition also allows $n < 0$ in~\eqref{singf} for certain $g$. One such example is the hyperscaling violating solution~\cite{Gouteraux:2011ce,Ogawa:2011bz,lizaetal,Shaghoulian:2011aa,Dong:2012se} (at $T=0$), where the metric functions have the scaling form 
\be \label{hsvs}
g(z)=b z^m \qquad f(z)=a z^n, \qquad z \to \infty  \ . 
\ee
We will  discuss the black hole case in section~\ref{sec:bh}. 



\subsection{Holographic Entanglement entropy: strip} \label{sec:eest}


We  first discuss the holographic entanglement entropy of the strip region~\eqref{defstrip}. It is obtained by minimizing the action:
\be 
S_{\rm strip} (R)= {L^{d-1} \ov 4 G_N} \ell^{d-2} A  
\ee
where $G_N$ is the bulk Newton constant and $A$ is the area functional~ \cite{Ryu:2006bv,Lewkowycz:2013nqa}.
If the spacetime is singular, as in the case of~\eqref{singf}, the minimal surface can become disconnected. In this case, the minimal surface consists of two disconnected straight planes $x (z)=\pm R $. The minimal surface area is independent from $R$ due to the translational symmetry of the problem. If the surface is connected, its area is given by
\be \label{act1} 
A  =  \int_{-R}^{R} d x \, {1 \ov z^{d-1}} \sqrt{1+ {z'^2 \ov f (z)}}   \ .
\ee
The shape of the entangling surface is specified by the boundary conditions
\be 
z(x=R) =0, \qquad z'(x=0) = 0 \ .
\ee
Since the action has no implicit dependence on $x$, we have an associated conserved quantity: 
\be 
 {1 \ov z^{d-1}} {1 \ov \sqrt{1 + {z'^2 \ov f}}} = {\rm const} \ .
\ee
This reduces the equation of motion to first order: 
\be \label{eomstr}
z' = -{1 \ov z^{d-1}} \sqrt{f (z) \le(z_t^{2 (d-1)}- z^{2 (d-1)}\ri)}\ ,
\ee
where $z_t = z(x=0)$ gives the tip of the minimal surface. $z_t$ is determined by requiring $z(R) = 0$. i.e.  
\be
{R}=\int_{0}^{z_t} du \  {u^{d-1} \ov \sqrt{f(u) \le(z_t^{2(d-1)}-u^{2(d-1)}\ri)}}=z_t \int_{0}^{1} dv \ {v^{d-1} \ov \sqrt{f(z_t v) (1-v^{2(d-1)})}} \ . \label{z0R}
\ee
Inverting this implicit equation gives the relation $z_t(R)$.
Using~\eqref{eomstr} we can also write~\eqref{act1} as 
\be 
A = {2 \ov z_t^{d-2}} \int_{{\de \ov z_t}}^1 {dv \ov v^{d-1}} {1  \ov \sqrt{f(z_t v) (1-v^{2(d-1)})}} \ 
\ee
where $\de$ is a UV cutoff. 

Expanding~\eqref{eomstr} near the boundary $z=0$, we find the expansion
\be\label{uvexp}
x(z)={R}-{z^d \ov d \, z_t^{d-1}}+\cdots\ . 
\ee
Varying~\eqref{act1} with respect to $R$ and using~\eqref{uvexp}, we find that
\be 
R^{d-1} {dA \ov dR} =2  \le({R \ov z_t}\ri)^{d-1} 
\ee
which implies that $\sR_d (R)$ defined in~\eqref{scke} is given by 
\be
\sR_d=  {L^{d-1} \ov  2G_N} \, \le({R \ov z_t}\ri)^{d-1} \ . \label{s2}
\ee
Thus to find $\sR_d$ it is enough to invert~\eqref{z0R} to obtain $z_t (R)$. \eqref{s2} was obtained before in~\cite{Myers:2012ed}.

\subsection{Holographic Entanglement entropy: sphere}  \label{sec:eesp}


Writing $d \vec x^2 = d \rho^2+ \rho^2 d \Om^2_{d-2} $ in polar coordinates, 
the entanglement entropy for a spherical region of radius $R$ can be written as 
\be 
S (R)= {L^{d-1} \ov 4 G_N} \om_{d-2} A \equiv K A \ , 
\ee
where $\om_{d-2}$ is the area of a unit $(d-2)$-dimensional sphere and $A$ is obtained by minimizing  the surface area
 \be \label{sphact}
A  =  \int_0^R d \rho \, {\rho^{d-2} \ov z^{d-1}} \sqrt{1 + {z'^2 \ov f (z)}} 
=  \int_0^{z_t} d z \, {\rho^{d-2} \ov z^{d-1}} \sqrt{\rho'^2 + {1 \ov f (z)}} \ ,
\ee
where $z_t$ denotes the tip of the minimal surface. The boundary conditions are 
\be \label{imbd}
\rho (z=0) = R , \qquad \rho (z_t) = 0, \qquad \rho'(z_t) = \infty  \ .
\ee
As discussed in~\cite{Liu:2012eea}, for~\eqref{singf} it is also possible for the minimal surface to have the cylinder topology, for which $z_t = \infty$ and the IR boundary 
conditions become
\be \label{bd2}
\rho (z) \to \rho_0 , \quad \rho'(z) \to 0, \qquad z \to \infty \ ,
\ee
with $\rho_0$ a finite constant.  The equation of motion can be written as 
\be 
(d-2)  {1 \ov f }  + (d-1) {\rho \rho'  \ov z} 
= {\rho } \sqrt{\rho'^2 + {1 \ov f}} \p_z \le( {\rho' \ov \sqrt{\rho'^2 + {1 \ov f}}} \ri) \label{rhoeom}
\ee
or 
\be \label{eom3}
 f z''  + \le({d-2 \ov \rho} z' + {(d-1) f \ov z} \ri) \le(f + z'^2  \ri)  - {\p_z f  \ov 2 } z'^2  = 0 \ .
\ee

In general, $\rho (z)$ can be expanded near the boundary in small $z$ as 
\be  \label{imexp}
\rho(z) = R  - {z^2 \ov 2R} + \cdots + c_d (R) z^d + \cdots  \ ,
\ee
where all coefficients except $c_d (R)$ can be determined locally (or in terms of $c_d$). One can show that~\cite{Liu:2012eea} 
\be \label{eopr}
{dA \ov dR} =  - d R^{d-2} c_d (R) - {e_d \ov R} + \cdots \ ,
 \ee
where $\cdots$ denotes non-universal terms which drop out when acted on with the 
differential operator in~\eqref{Scen}, and $e_d$ is a constant, which is nonvanishing only for $d=4,8, \cdots$.
 Using~\eqref{eopr} one can express the REE~\eqref{Scen} in terms of $c_d (R)$. For example, 
for $d=3$
\be \label{rr1}
{1 \ov K} \sS_3 (R)= - 3 R^2 c_3 (R) + 3 \int_0^R dR \, R c_3 (R)  + C 
\ee
where $C$ is determined by requiring that $\sS_3 (R=0)$ reduces to the value at the UV fixed point,
and  for $d=4$, 
\be \label{rr2}
{1 \ov  K} \sS_4 = 1  -  2 R^3 c_4 (R) -   2 R^4 {d c_4 \ov dR}  \ . 
\ee

 One could also obtain $\sS_d$ by directly evaluating the action~\eqref{sphact} and then taking the appropriate derivatives~\eqref{Scen}.

\subsection{Strategy for obtaining the entanglement entropy for a sphere}


In general it is not possible to solve~\eqref{rhoeom} or~\eqref{eom3} exactly. Here we outline  
a strategy to obtain the large $R$ expansion of $S(R)$ (or $\sS_d (R)$) via a matching procedure:
\ben
\item Expand $\rho (R)$ in~\eqref{rhoeom} in $1/R$ as
\be 
\rho(z)= R - {\rho_1(z) \ov R } - {\rho_3 (z) \ov R^3 }+ \dots -{\hat\rho(z)\ov  R^{\nu}}+\dots \ .  \label{exp}
\ee
Note that the above expansion applies to the vacuum. For a black hole geometry 
one should include all integer powers of $1/R$ as we will discuss in more detail in section~\ref{sec:bh}.
The expansion~\eqref{exp} should be considered as an ansatz, motived by~\eqref{irex} one wants to show, but should be ultimately confirmed by the mathematical consistency of the expansion itself (and the matching 
described below).

Depending on the IR behavior of a system, the large $R$ expansion~\eqref{exp} can contain 
terms which are not odd powers of $1/R$.  We have denoted the exponent of the first such term in~\eqref{exp} as 
$\nu$, whose value will be determined later. The expansion is valid for $\rho (z)$ close to $R$, i.e. ${\rho_1 \ov R} \ll R$ etc. It is clearly valid near the boundary (i.e. small $z$ where~\eqref{imexp} applies), but depending on the configuration of the minimal surface it may 
also apply to regions, where $z$ is not small, as far as higher order terms in~\eqref{exp} remain 
small compared to $R$. 

 \item Determine the IR part (i.e. in the region where~\eqref{pep1} or~\eqref{singf} applies) of the minimal surface in a large $R$ expansion. 
This has to be done  case by case, as the IR expansions are different for different IR geometries. 
 
\item Match the two solutions in the appropriate matching region.
At the end of the matching procedure all free constants get determined including $c_d (R)$ of~\eqref{imexp}. 

\begin{figure}[!h]
\begin{center}
\includegraphics[scale=0.33]{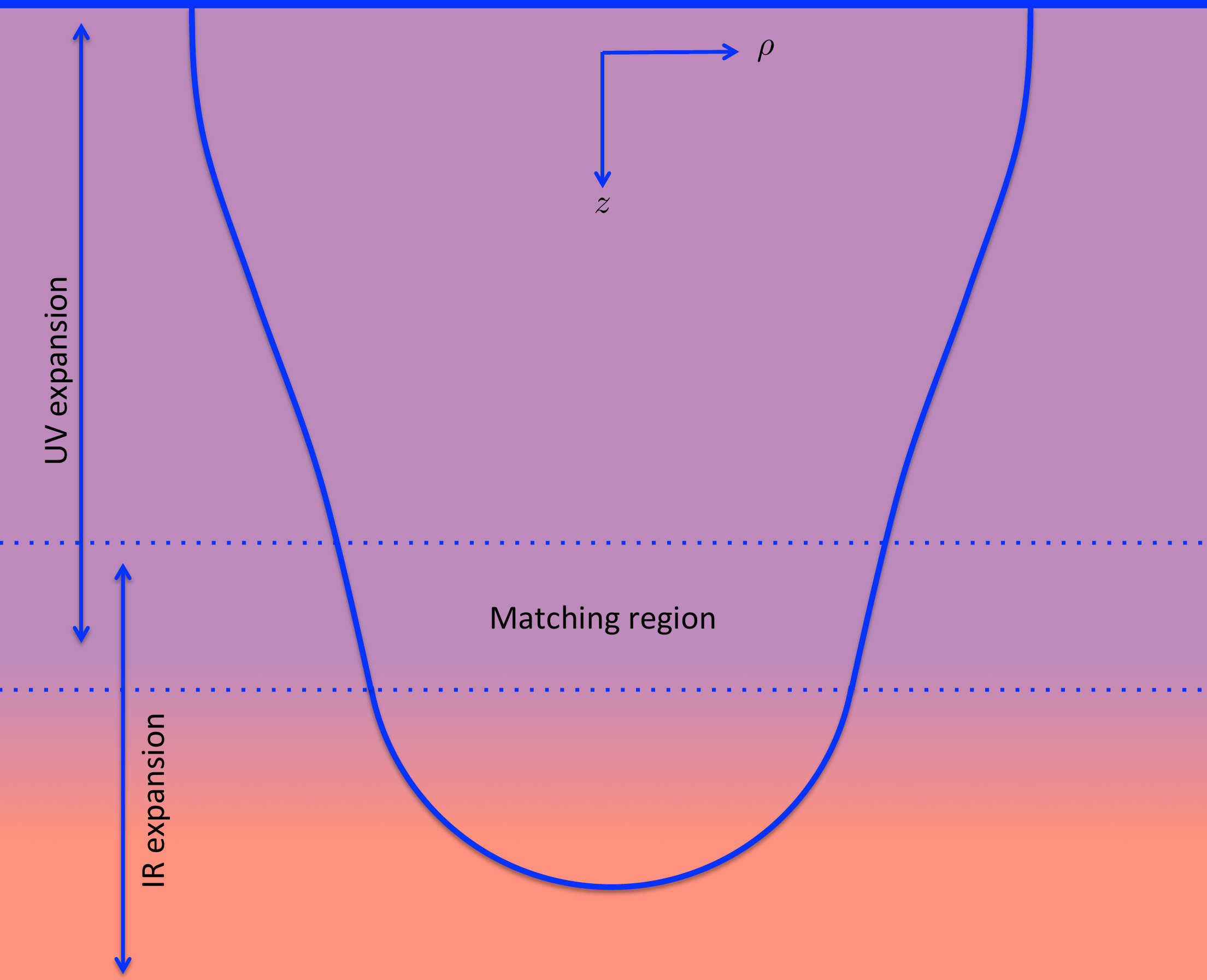}
\end{center}
\caption{Sketch of the $R\to\infty$ minimal surface in  a domain wall geometry~\eqref{1}--\eqref{pep1}. The violet and red regions represent  the UV and IR  regions of the expansion. The UV and IR solutions overlap in the matching region, which is used to determine the parameters of the two expansions. 
}\label{explain}
\end{figure}

\begin{figure}[!h]
\subfigure[Minimal surface for a gapped geometry~\eqref{singf} with $n > 2$.  
\label{qualitativea}]{\includegraphics[scale=0.7]{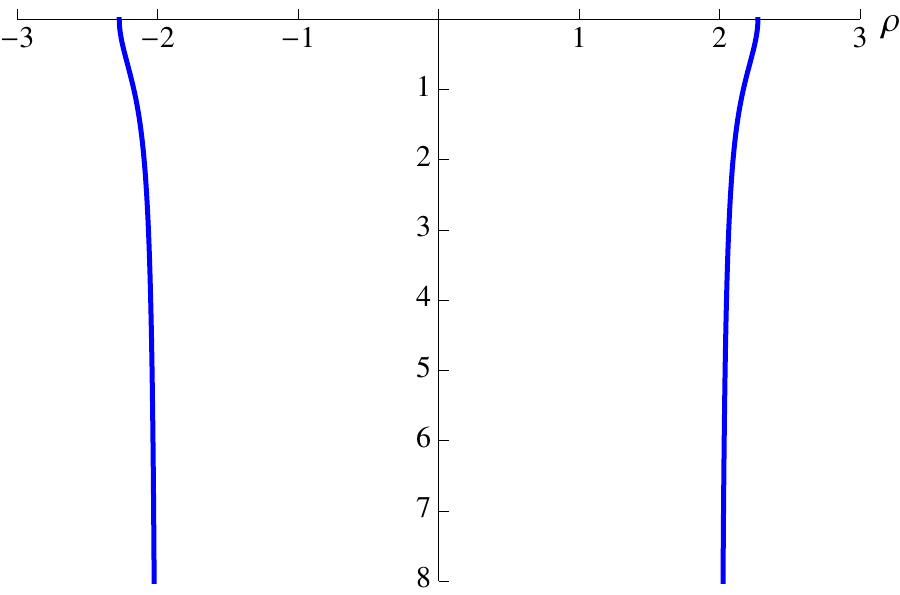}}\hspace{1cm}
\subfigure[Minimal surface for a scaling geometry with $0< n <2$. 
\label{qualitativeb}]{\includegraphics[scale=0.7]{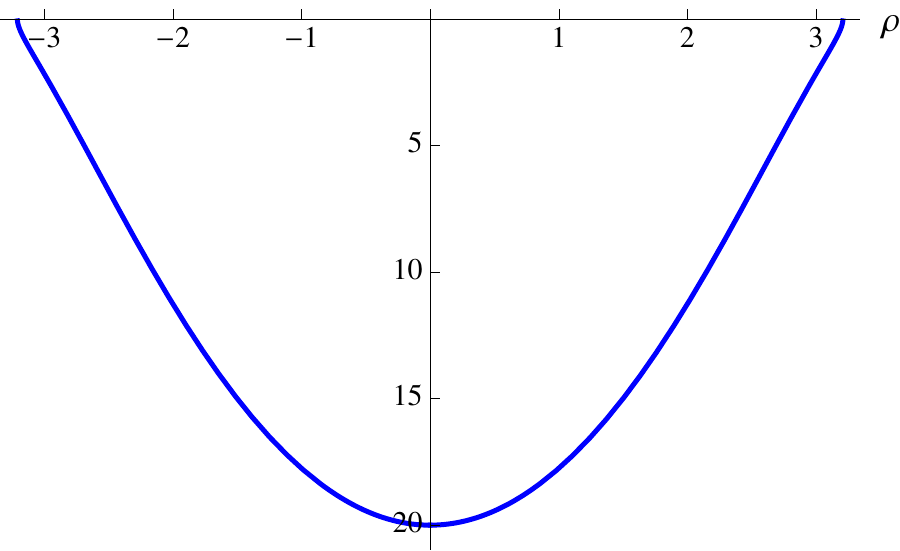}}
\subfigure[Minimal surface for a domain wall geometry with IR geometry given by~\eqref{pep1}. 
]{\includegraphics[scale=0.7]{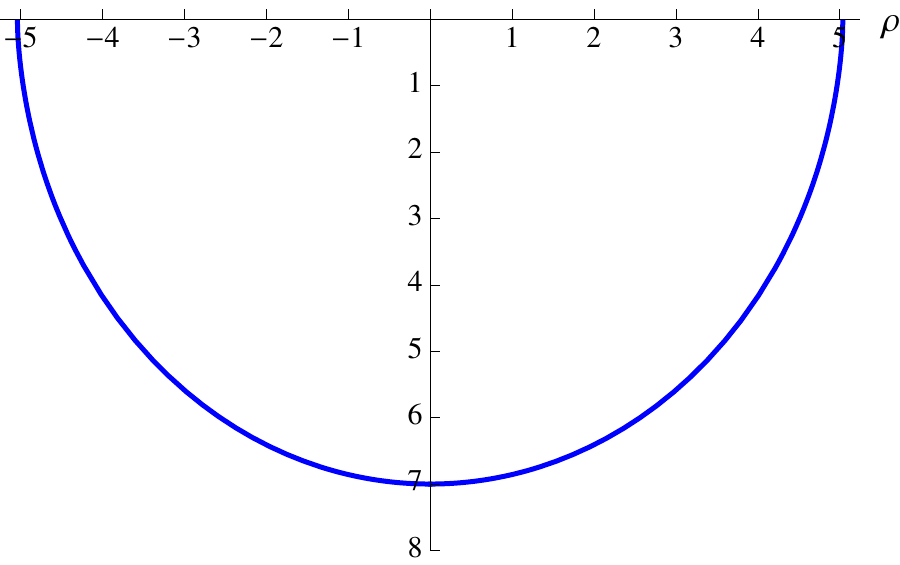}}\hspace{1cm}
\subfigure[Minimal surface for a Schwarzshild black hole.  
The beyond the horizon region is marked by gray.]{\includegraphics[scale=0.7]{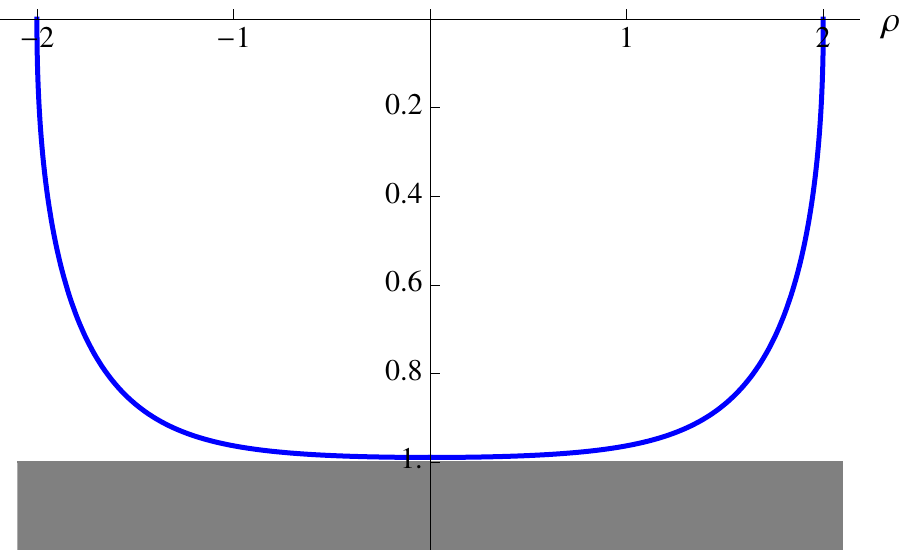}}
\caption{Samples of minimal surfaces for IR geometries that fall into four different categories. 
} 
\label{qualitative}
\end{figure}

\een
See Fig.~\ref{explain} for an illustration of the matching procedure and in Fig.~\ref{qualitative} we show how the minimal surfaces look for different IR  geometries. 

From~\eqref{exp} we see that $c_d(R)$ in~\eqref{imexp} takes the following expansion 
\be \label{uwme}
c_d(R)=-{b_1\ov d R}-{b_3 \ov d R^3}+\dots -{\hat b\ov d R^{\nu}}+\cdots\ ,
\ee
where $b_n$ and $\hat b$ are some $R$-independent constants. 
It follows from~\eqref{eopr} and~\eqref{Scen} that a term proportional to $1/R^{n}$ 
in~\eqref{uwme} contributes to $\sS_d$ a term of order $1/R^{n-d+1}$,  whose coefficient 
contains a factor $(n-1) (n-3) \cdots (n- (d-2))$ for odd $d$, or $(n-1) (n-3) \cdots (n- (d-3))$ for even $d$.
 Thus, among the integer powers of $1/R$
in~\eqref{uwme}, in odd $d$ the first possible nonvanishing contribution to $\sS_d$ comes from $b_{d}$
giving a term proportional to $1/R$, and in even $d$ the first possible nonvanishing contribution comes from $b_{d-1}$ giving a term of order $O(R^0)$. Furthermore, the terms in~\eqref{uwme} with odd integer powers will only give rise to  odd inverse powers of $R$ in odd dimensions and even inverse powers in even dimensions, as in the second line of~\eqref{irex}.  
Finally from~\eqref{eopr} and~\eqref{Scen}, a term proportional to $R^{-\nu}$ in~\eqref{uwme} gives a contribution 
\be \label{eooe}
\sS_d (R) = \cdots + R^{d-1-\nu}\,{K \, \hat b \ov (d-1 - \nu)\, \Ga ({1-\nu \ov 2})\,  \Ga({d \ov 2})}  
\bca
        \sqrt{\pi}\, \Ga ({d-\nu \ov 2})   & {\rm odd} \; d \cr
         2\, \Ga ({d+1 - \nu \ov 2}) & {\rm even} \; d
         \eca +   
\cdots  \ .
\ee

\subsection{UV expansion}\label{sec:uvexp}


We now examine more explicitly the UV expansion~\eqref{exp} for the sphere, which is the same for all geometries of the form~\eqref{1}. The IR expansion and matching will be discussed in later sections case by case. 

The equation for $\rho_i (z)$ can be written as 
\be  \label{ehor}
{z^{d-1} \ov \sqrt{f}} \le({\sqrt{f} \ov z^{d-1}} \rho_i' \ri)' = s_i \ ,
\ee
where $s_i$ denotes a source from lower order terms with, for example, 
\be 
s_1 = - {d-2 \ov f} \ .
\ee
The equation for $\rho_1$ can be readily integrated to give
\be \label{rho1s}
\rho_1(z)= b_1 \rho_{hom}(z) - (d-2)
\int_{0}^z du \ {u^{d-1}\ov \sqrt{f(u)}} \int_{\infty}^u dv \ {1 \ov v^{d-1}\sqrt{f(v)}}  \ , 
\ee 
where $b_1$ is an integration constant 
and    
$\rho_{hom}$ is the homogenous solution to~\eqref{ehor}
\be 
\rho_{hom}(z)=\int_{0}^z du \ {u^{d-1}\ov \sqrt{f(u)}} \ .
\ee
In particular because its unique $R$-dependence there are no source terms for $\hat\rho(z)$, thus it takes the form:
\be \label{rhoha}
\hat\rho(z)=\hat b  \rho_{hom}(z) \ .
\ee
As $z \to 0$, $\rho_1$ and $\hat \rho$ has the leading behavior  (for $d \geq 2$)
\be  \label{eorn}
\rho_1 (z) =
 O(z^2) , \qquad  \hat \rho (z) = {\hat b \ov d} z^d + \cdots \ . 
\ee
Note that the normalization of $ \rho_{hom}$ in~\eqref{rhoha} was chosen such that the contribution to $c_d(R)$, read off from~\eqref{eorn}, gives the term appearing in~\eqref{uwme}.


\section{Gapped and scaling geometries}\label{sec:gapped}

In this section we consider the large $R$ behavior of the REE for holographic systems,
whose IR geometry is described by~\eqref{singf}. As mentioned below~\eqref{singf} 
there is an important difference between $n > 2$ and $ n \leq 2$, to which we refer as gapped and scaling geometries 
respectively. For comparison we will treat them side by side. 
We will first consider the strip and then the sphere case.







\subsection{Strip}

In~\eqref{z0R} to leading order in large $z_t$, we can replace $f(z)$ in the integrand by its large $z$ behavior $f (z) =a z^n$, leading to 
\be \label{rzte}
R (z_t)  = z_t \le[\int_{0}^{1} dv \ {v^{d-1} \ov \sqrt{a (z_t v)^n(1-v^{2(d-1)})}}+\cdots\ri]
= {\al \ov \sqrt{a}} z_t^{1-{n\ov 2}}
+ \cdots , \quad z_t \to \infty
\ee
with 
\be \label{etaf}
\al =  {2\sqrt{\pi}\, \Gamma\le(\ha + {\eta\ov 4}\ri)\ov\le(2-n\ri)\Gamma\le(\eta \ov 4\ri)}, \qquad
\eta \equiv {2-n \ov d-1} \ . 
\ee
For small $z_t$ we can replace $f (z_t v)$ in~\eqref{z0R} by $1$ and thus 
\be 
R (z_t ) = {z_t \ov d} + \cdots , \qquad z_t \to 0 \ .
\ee

For $n > 2$, the function $R(z_t)$ then 
goes to zero for both $z_t \to 0$ and $z_t \to \infty$, and thus 
must have a maximum in between at some $z_t^{(max)}$.  Introducing 
\be
R_{max}=z_t^{(max)} \int_{0}^{1} dv \ {v^{d-1} \ov \sqrt{f(z_t^{(max)} v) (1-v^{2(d-1)})}} 
\ee
we conclude that for $R>R_{max}$ there is no minimal surface with strip topology. 
Instead, the minimal surface is just two disconnected straight planes $x (z)=\pm R $. The minimal surface area is independent from $R$ due to the translational symmetry of the problem. We conclude that for $n > 2$  in the $R\to \infty$ limit $S$ becomes independent of $R$, hence $\sR_d(R>R_{max})=0$.
For $n =2$, $R (z_t) \to {\rm const}$ at large $z_t$, and again in this case there is no minimal surface of strip topology and $\sR_d  (R > R_{max}) = 0$. 

For $n < 2$, inserting~\eqref{rzte} into~\eqref{s2} we find that 
\be \label{stripan}
\sR_d={L^{d-1} \ov 2 G_N} 
\le({\al^2 \ov a R^n} \ri)^{1 \ov \eta} + \cdots  \propto R^{-\beta} \ , \quad 0 < n < 2 \ 
\ee
with 
\be\label{defbe}
\beta = n {d-1 \ov 2-n} =  {n \ov \eta} \ .
\ee
This result also applies to a hyperscaling violating geometry~\eqref{hsvs}, 
and agrees with the scaling derived in~\cite{Dong:2012se}. 

\subsection{Sphere}

Since for $d=2$, the sphere and strip coincide (the answer is then given by~\eqref{stripan}),  we will restrict our discussion below to $d \geq 3$.

\subsubsection{IR expansion}\label{sec:scalingIR} 

We first consider the behavior of the minimal surface in the IR geometry~\eqref{singf}.
Plugging $ f(z) = a z^n$ into~\eqref{rhoeom} we notice that if $\bar \rho (z)$ satisfies
the resulting equation with $a=1$, then  
\be \label{aEqOne}
\rho (z) =  \bar \rho  \le(a^{-{1 \ov 2-n}} z\ri) 
\ee
satisfies~\eqref{rhoeom} for any $a$. Furthermore, 
equation~\eqref{rhoeom} is invariant under the scaling
\be 
\rho \to \lam^{2-n\ov 2} \rho \ , \qquad z \to \lam z \ ,
\ee
which implies that 
if $\rho(z)$ is a solution to~\eqref{rhoeom}, so is $\rho_\lam (z) = \lam^{2-n \ov 2} \rho(\lam^{-1} z)$.

Solutions of two different topologies are possible. As discussed in~\cite{Liu:2012eea}, 
for  $n > 2$, in the large $R$ limit the minimal surface has the topology of a cylinder, 
while for $n \leq 2$, the minimal surface has the topology of a disk. See Fig.~\ref{qualitativea} and  
Fig.~\ref{qualitativeb}. 


For a solution of cylinder topology (i.e. for $n > 2$) the IR solution  
satisfies 
\be 
\rho (z) \to  \rho_0 , \quad \Rightarrow \quad  \rho_\lam (z)  \to \lam^{2-n \ov 2} \rho_0 , \qquad z \to \infty  \ .
\ee
Introducing a solution $\bar \rho_c (z)$ to~\eqref{rhoeom} with $a=1$,  which satisfies the condition 
\be 
\bar \rho_c (z \to \infty) = 1 \ ,
\ee
we can write a general $\rho (z)$ in a scaling form 
\be 
\rho (z) = \rho_0 \bar \rho_c (v), \qquad v \equiv (\rho_0^2 a)^{1 \ov n-2}  z \ .
\ee
From~\eqref{rhoeom},  $\rho (z)$ has the large $z$ expansion (see also Appendix C of~\cite{Liu:2012eea}) 
\be \label{ng2s}
\rho (z) = \rho_0+ {2 (d-2)   \ov  \rho_0 a (n-2) (n+2d-4)}  z^{2-n} + \cdots , \quad z \to \infty, \quad n > 2 \ .
 \ee

For a solution of disk topology (i.e. for $n \leq 2$),   there should exist a $z_t < \infty$, where
\be
z_t = z (\rho =0) \quad {\rm or} \quad \rho (z_t) = 0 \ . 
\ee
Now introducing a solution $\bar \rho_d  (z)$ to~\eqref{rhoeom} with $a=1$, which 
satisfies the boundary condition $\bar \rho_d (1) = 0$, 
we can write 
$\rho (z)$ in a scaling form
\be \label{jep}
\rho (z) = {z_t^{(2-n)/2}\ov \sqrt{a}} \, \bar{\rho}_d \le(u \ri), \quad {\rm with} \quad
u \equiv {z\ov  z_t}, \quad  \bar \rho_d (u=1) = 0 \ .
\ee
Note that by taking $z_t$ sufficiently large, $u$ can be small even for $z \gg z_{CO}$, 
where~\eqref{singf} applies.  
Expanding $\bar \rho_d$ in {\it small} $u$ one finds that 
\bea
\bar{\rho}_d (u)&=&\bar \al_0+{\al_1\ov \bar{\al}_0} u^{2-n}+{\al_2\ov \bar{\al}_0^3} u^{2(2-n)}+\cdots \cr
&+ & {\bar{h} \ov \bar{\al}_0^{2 \ov \eta}}\, u^{d-n/2}+\cdots\ , \quad u \to 0 \ , \label{jepexp}
\eea
where $\eta$ was introduced in~\eqref{etaf} and
\be
\al_1=-{2(d-2)\ov (2-n)(2d-4+n)}\ ,  \quad \cdots  \ . \label{a1}
\ee
 $\bar \al_0$ and $\bar{h}$ are numerical constants that can be obtained by numerically solving the equation of motion.   Using~\eqref{jepexp}, we then get the expansion for $\rho(z)$:
\bea
\rho(z)&=&\al_0+{\al_1\ov a \al_0} z^{2-n}+ {\al_2\ov a^2\al_0^3 } z^{2 (2-n)} +
\cdots \cr
&+ & {\bar{h} \ov a^{{1 \ov \eta} + \ha} \al_0^{2\ov \eta}}\, z^{d-n/2}+\cdots \label{scalingirexp}
\eea
with  
\be \label{893}
\al_0 \equiv   {\bar{\al}_0\, z_t^{(2-n)/2}\ov \sqrt{a}} \ .
\ee
 This is all the information we need about the IR solution.
Note that the above expansion applies to the range of $z$, which satisfies
\be \label{ovp1}
z \gg z_{CO}, \qquad  {z\ov z_t} \ll 1 \ .
\ee
The small $u$ expansion~\eqref{jepexp} is singular for $n=2$, as can be seen from~\eqref{a1}. Hence the $n=2$ case should be treated separately, see Appendix~\ref{app:n2}.

\subsubsection{Matching} 

We first examine the UV solutions~\eqref{rho1s} and~\eqref{rhoha} for a sufficiently large $z$ so that~\eqref{singf} applies. At leading order in large $z$, we then find that 
\bea
\rho_1(z)&=&\int_{0}^z du \ {u^{d-1}\ov \sqrt{f(u)}} \le(b_1 +(d-2)\int_{u}^\infty dv \ {1 \ov v^{d-1}\sqrt{f(v)}}\ri)\cr
&=&{b_1 \ov \sqrt{a}}\, {z^{d-n/2}\ov d-n/2}\le(1+\dots\ri)+{2(d-2)\ov (2-n)(2d-4+n)\,a} \, z^{2-n}\le(1+\dots\ri)
\label{rhi1}
\\
\hat\rho(z)&=&  
{\hat b\ov \sqrt{a}}\, {z^{d-n/2}\ov d-n/2}\le(1+\dots\ri) \ .
\label{gje1}
\eea

Plugging~\eqref{rhi1} and~\eqref{gje1} into~\eqref{exp}, we see that to match the UV expansion with the $n > 2$ solution~\eqref{ng2s} at large $z$, we require 
\be\label{ScalingMatch}
b_1 = \hat b = 0, \qquad \rho_0 = R \ . 
\ee
We see that the UV expansion in fact directly matches to the behavior at $z \to \infty$ without the need of an intermediate matching region. 
Thus in this case the UV expansion~\eqref{exp} can be extended to arbitrary $z$ without  breaking down, which  
can be verified by showing that higher order terms are all finite for any $z$. This is also intuitively clear from Fig.~\ref{qualitativea} where for large $R$ the minimal surface has a large radius at any $z$. Note that, since $\hat b =0$, the non-integer $\nu$ term in~\eqref{exp} is not present.

For $n < 2$, where the minimal surface has the topology of a disk, the UV expansion is destined to break down at certain point before the tip of the minimal surface is reached. In the region~\eqref{ovp1} both the IR and UV expansions apply, and by comparing~\eqref{rhi1} and~\eqref{gje1}  
with~\eqref{scalingirexp}, we find that they match precisely provided that
\be \label{unep}
\al_0 = R, \qquad b_1 =0, \qquad \hat b = - \le(d-{n \ov 2}\ri) \bar h a^{-{1 \ov \eta}}, \qquad \nu = {2 \ov \eta} \ .
\ee 
From~\eqref{893} we conclude that $z_t$ scales with $R$ as 
\be \label{894}
z_t \sim R^{2 \ov 2-n}, \qquad R \to \infty \ . 
\ee

Again, the story for $n=2$ is discussed in Appendix~\ref{app:n2} with equation~\eqref{894}  replaced 
by 
\be 
z_t \sim \exp \le(-{(d-1)^2 \, a\ov 2(d-2)}\, R^2\ri) \ .
\ee

\subsubsection{Asymptotic expansion of the REE}

We will now obtain the leading order behavior of the REE in the large $R$ limit. 

\medskip

\noindent {\bf  1. $n > 2$}

\medskip

Let us first consider $n > 2$. From the discussion below~\eqref{uwme}, we expect the leading order term for odd 
$d$ to be proportional to $1/R$, which comes from the $1/R^{d}$ term in the expansion of $c_d$. 
For even $d$, the leading term can in principle be $1/R^0$, which comes from the $1/R^{d-1}$ term in the expansion of $c_d$. Note, however, since this a gapped system, we expect the order $1/R^0$ term to vanish. So, for even $d$, the leading term should come from the $1/R^{d+1}$ term. 

Since even for $d=3$ we would need to know $c_3(R)$ to $1/R^3$ order, and we only worked out $\rho_1$ (which only determines $c_3(R)$ to $1/R$), our results seem insufficient to determine the $1/R$ contribution to $\sS_3$. 
However, the $1/R$ contribution to $\sS_3$ can be obtained by directly evaluating the on-shell action~\cite{Klebanov:2012yf}, as the $1/R$ piece is the next to leading term in the large $R$ expansion of $S$. 
For $d=4$, we can use $\rho_1$ to verify that the $1/R^0$ term (in the REE) vanishes as expected for a gapped system.
With due diligence, it is straightforward to work out higher order terms, but will not be attempted here. 

For $d=3$, plugging~\eqref{exp} into~\eqref{sphact} we have the expansion 
\bea
A&=&R\,\int_\delta^\infty dz\, {1\ov z^2\,\sqrt{f(z)}}+{1\ov R} \int_0^{\infty} dz \ \le[{\sqrt{f(z)}\ov 2 z^2} \rho_1'(z)^2 - {\rho_1(z)\ov  z^2\, \sqrt{f(z)}} \ri]+O\le({1\ov R^3}\ri)  \cr
& = & \#\, R+{1\ov R} \int_0^{\infty} dz \ \le[{\sqrt{f(z)}\ov 2 z^2} \rho_1'(z)^2 + \rho_1(z) \le({\sqrt{f}\rho_1' \ov z^2}\ri)'  \ri]+O\le({1\ov R^3}\ri) \ , \label{AOnShell1}
\eea
where in the second line we have used~\eqref{ehor}. 
Integrating by parts the second term in the integrand we find that
\be \label{AOnShell2}
A = \# R - {1 \ov R} \int_0^{\infty} dz \ {\sqrt{f(z)}\ov  2z^2} \rho_1'(z)^2 \ ,  
\ee
where the boundary terms vanishe due to~\eqref{eorn} and~\eqref{ng2s}.
We thus find that
\be \label{a1e}
A=\#\, R- {a_1 \ov 2R} +\cdots , \quad
a_1 =
\int_0^{\infty} dz \ {z^2 \ov \sqrt{f(z)}}  \le[\int_z^\infty dv \ {1\ov v^2 \sqrt{f(v)}}\ri]^2 \ .
\ee
It is desirable to make work with dimensionless coefficients that only depend on ratios of scales. We can use
\be\label{ascale}
\tilde\mu\equiv a^{1/n} 
\ee
as an energy scale and define the dimensionless coefficient
\be
s_1\equiv \tilde\mu \, a_1=\int_0^{\infty} dz \ {z^2 \ov \sqrt{f(z/\tilde\mu)}}  \le[\int_z^\infty dv \ {1\ov v^2 \sqrt{f(v/\tilde\mu)}}\ri]^2 \ ,
\ee
where all integration variables are dimensionless, and $s_1$ only depends on ratios of scales, e.g.~$(\tilde\mu\, z_{CO})$.
Finally, we obtain
\be \label{ss0}
\sS_3={s_1 K \ov \tilde\mu R}+\cdots , \qquad n > 2 \ .
\ee
This result agrees with those in~\cite{Klebanov:2012yf}. It is interesting to note that the coefficient of $1/R$ term depends on the full spacetime metric, i.e. 
in terms of the boundary theory, the full RG trajectory. 

For $d=4$, the expansion of $A$ has the form 
\be \label{a4}
A = a_0 R^2 + a_2 + O(1/R^2) 
\ee
where
\be
a_0 = \int_\delta^\infty dz\, {1\ov z^3\,\sqrt{f(z)}},  \qquad a_2=  \int_\de^\infty dz\, {f(z) \rho_1'(z)^2- 4 \rho_1(z)\ov 2z^3\,\sqrt{f(z)}} 
\ee
with $\delta$ a UV cutoff. Neither of the first two terms indicated in~\eqref{a4} will contribute to $\sS_4$ after differentiations in~\eqref{Scen}.  As expected, $a_0 \sim1/ \de^2$ is UV divergent. 
$a_2$ contains a logarithmic UV divergence $\log {\de \mu}$, where $\mu$ is mass scale controlling the leading 
relevant perturbation from the UV fixed point.  
At large $z$, from~\eqref{rhi1} and~\eqref{ScalingMatch}  $\rho_1\sim z^{2-n}$, hence the integrand for
$a_2$ goes as $\sim z^{-1-3n/2}$, and the integral is convergent at the IR end. 
An IR divergent $a_2$ would signal a possible $\log R$ term. Thus we conclude that the leading order contribution for $d=4$ is of order $1/R^2$, consistent with our expectation that the system is gapped.

\medskip 

\noindent {\bf 2. $n \leq 2$}

\medskip

For $n < 2$, $\hat b$ in~\eqref{uwme} is nonzero and its contribution to $\sS_d$ can be directly 
written down from~\eqref{unep}
\be \label{ss1}
\sS_d =  e_n  {K \ov a^{1 \ov \eta} R^\beta}+ \bca 
         O(R^{-1}) & d \; {\rm odd} \cr
         O(R^{-2}) & d \; {\rm even}
       \eca \ ,
\ee
where $\eta$ and $\beta$ were defined in~\eqref{etaf} and~\eqref{defbe} respectively, and 
\be \label{defen}
e_n =  {d - n/2 \ov n}   {\eta \, \bar h \ov \Gamma\le(d\ov2\ri)\Gamma\le(\ha - {1 \ov \eta}\ri)}\times 
\begin{cases}
\sqrt{\pi}  \Gamma\le(\ha - {n \ov 2 \eta}\ri)  & \text{$d$ odd}\\
 2 \Gamma\le(1 - {n \ov 2 \eta}\ri) &  \text{$d$ even}
\end{cases} \ .
\ee
For $n=2$ the first term in~\eqref{ss1} should be replaced by (see~\eqref{n2nonanal} and Appendix~\ref{app:n2})
\be\label{expsup}
 \sS_d^{\text{(non-analytic)}}\propto \le(a\,R^2\ri)^{t}\, \exp\le(-{(d-1)^2 \, a\ov 2(d-2)}\, R^2\ri) \qquad t\equiv{d-3\ov 2}+\le[d\ov 2\ri] \ .
 \ee

\begin{figure}[h!]
\begin{center}
\includegraphics[scale=0.6]{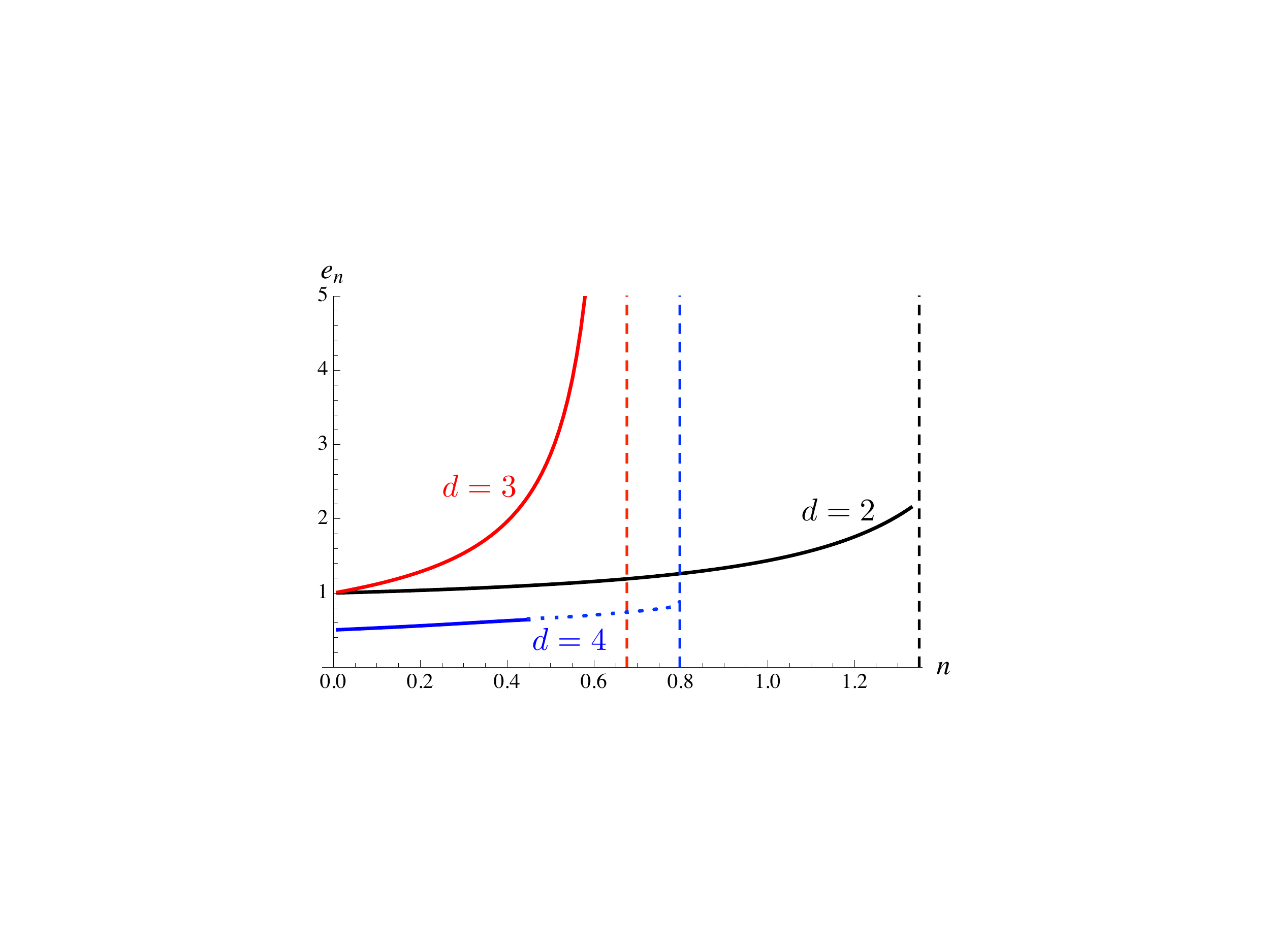}
\caption{$e_n$ plotted as a function of $n$ for $d=2,3$ and $4$. The vertical dashed lines indicate $n_c$. \eqref{defen} consists of numerical factors and $\bar{h}$, which is a constant determined by the IR solution, $\bar{\rho}_d$.  $\bar{h}$ was obtained by numerically determining $\bar{\rho}_d$ and fitting the small $u$ expansion~\eqref{jepexp}. For $d=2$ we know the exact answer from~\eqref{stripan}; the data points lie exactly on the analytically determined curve. For $d=4$ the dotted part of the line is an extrapolation of the solid line; we do not have reliable numerical results in that region for $\bar{h}$.  \label{enplot} }
\end{center}
\end{figure}

Below for convenience we will refer to the first term in~\eqref{ss1} (or~\eqref{ss2}) as  ``non-analytic'', while terms of inverse odd powers in odd dimensions (and even inverse powers in even dimensions) as ``analytic.'' 
Note that the non-analytic term is the leading contribution in the large $R$ limit when
\be \label{leading}
n < n_c \equiv \bca {2 \ov d} & \text{$d$ odd} \cr
         {4 \ov d+1} & \text{$d$ even} 
         \eca \ ,
         \ee
in which case one can check that the coefficient $e_n$ is positive. In Fig.~\ref{enplot} we plotted $e_n$ for $d=2,3$ and $4$. Note that for odd $d$, $e_n$ diverges as $n\to n_c$, while for even $d$ it stays finite.\footnote{For $d=2$ apart from the numerical results, we can analyze the analytic answer given in~\eqref{stripan}.} Despite appearances the numerical factors multiplying $\bar{h}$ in~\eqref{defen} do not diverge at $n=n_c$, hence the features described in Fig.~\ref{enplot} are caused by $\bar{h}$.

Let us consider the $n \to n_c$ limit of~\eqref{ss1} for odd $d$. Because $e_n$ diverges as $n \to n_c$, in order for~\eqref{ss1}  to have a smooth limit,  we expect the coefficient of the $1/R$ term in~\eqref{ss1} to diverge too, in a way that the divergences cancel resulting in a logarithmic term
\be 
\sS_d = \# 
 {\log R \ov R}  + \cdots, \qquad n= {2 \ov d}, \quad \text{$d$ odd} \ .
\ee
The coefficient of the logarithmic term is given by the residue of~\eqref{defen} in the limit $n \to n_c$.  In contrast, for even $d$, 
$e_n$ is finite  at $ n = n_c$. Thus, the leading term will simply be of order $1/R^2$ with no
 logarithmic enhancement (there can still be logarithmic terms at higher orders). 

For $d=3$  one can calculate the coefficient of $1/R$ term in~\eqref{ss1} similar to $n > 2$ case discussed. 
See Appendix~\ref{app:scalingonshell} for a derivation.
One finds 
\be \label{ss2}
\sS_3 (R) = e_n  {K \ov (\tilde \mu R)^{2 n \ov 2-n}} + {K s_1 \ov \tilde\mu R} + \cdots \ ,
\ee
where  $s_1$ is given by~\eqref{a1e} for $n > {2 \ov 3}$, and for $n < {2 \ov 3}$ by 
\be\label{a12}
s_1=\int_0^{\infty} dz \  \le({z^2 \ov \sqrt{f(z/\tilde\mu)}}\le[\int_z^\infty dv \ {1\ov v^2 \sqrt{f(v/\tilde\mu)}}\ri]^2-{4 \ov (2 + n)^2}\, { 1\ov z^{3n/2}}\ri) \ . 
\ee
In this case, we can work out explicitly how the divergence in the limit $n \to n_c$ cancels between the coefficients of the analytic and non-analytic pieces. Note that the divergence in $s_1$ comes from the second term in the integrand in~\eqref{a12}\footnote{At first sight it seems puzzling that the divergence comes from the UV region, $z=0$. However, this is just an artifact of the subtraction we chose. }
\be
s_1=-{3\ov 8\, ( 2/3-n)}+\dots \ .
\ee
The numerical results presented in Fig.~\ref{enplot} are consistent with the behavior
\be
e_n={3\ov 8\, ( 2/3-n)}+\dots \ ,
\ee
to $1\%$ precision. Plugging into~\eqref{ss2} then gives 
\be \label{ss2n2}
\sS_3 (R) = K\, {27\ov 32}\,  {\log \tilde\mu R \ov \tilde\mu R} + {\# \ov  R} +\cdots\ .
\ee
We can perform the same calculation with $n=2/3$ fixed from the beginning, and we get the same result, see~\eqref{n23result2}.


\subsection{Discussion}

We now briefly summarize the results by comparing between the strip and the sphere, and between $n < 2$, $ n > 2$ and $n=2$ geometries.

 The presence of analytic terms for the sphere can be expected from the general structure of local contributions to the entanglement entropy~\cite{tarun,Liu:2012eea}, which implies the existence of  terms of the form $1/R + 1/R^3 + \cdots$ for odd dimensions and  $1/R^2 + 1/R^4 + \cdots$ for even dimensions. 
 Note  the coefficient~\eqref{a1e},~\eqref{a12} of the $1/R$ term in~\eqref{ss0} and~\eqref{ss2} depend on the full spacetime metric and  thus the full RG trajectory. This is consistent with the physical interpretation that such coefficients encode the contributions from degrees of freedom at all shorter length scales compared to $R$.\footnote{In~\eqref{a1e},~\eqref{a12} the upper limits of the integrals are $\infty$, as 
 we are considering $R \to \infty$ limit.}  For a strip, other than the area, all curvature invariants associated with the entangling surface vanish, and thus 
the analytic terms are altogether absent.

For $n <2$ geometries, non-analytic terms are present for both the strip and the sphere, and have the same scaling.  We note that the non-analytic terms (including the coefficients) are  solely determined by the IR geometry. From the boundary perspective they can be interpreted as being determined by the IR physics. 
The presence of these non-analytic terms (despite the fact that they could be subleading compared to analytic terms)
imply that the IR phase described by~\eqref{singf} is not fully gapped, and some IR gapless degrees of freedom 
are likely responsible for the non-analytic scaling behavior. For this reason we refer to such geometries as scaling geometries. Note that due to the singularity at $z = \infty$, we should view the region~\eqref{singf} as describing
an intermediate scaling regime. It likely does not describe the genuine IR phase, which depends on how the singularity is resolved. 
Thus our discussion above 
should be interpreted as giving the behavior of $\sS (R)$ for an intermediate regime. 
  We will see some explicit examples in the next section.

In contrast for $n > 2$, there is no non-analytic term and we expect the dual system to be fully gapped in the IR. 

For $n=2$ the strip and sphere entanglement entropies show different behaviors as emphasized recently by~\cite{Shaghoulian:2013qia}. For $R\to \infty$ the minimal surface for a strip is disconnected, and hence there is no non-analytic term in the expansion of $\sR_d$. However, for a spherical entangling surface the topology of the minimal surface is a disc, and $\sS_d$ contains an exponentially small term~\eqref{expsup}. 
In next section, by examining the spectral function of a scalar operator, we argue that an $n=2$ geometry describes a gapped phase, but with a continuous spectrum above the gap. 

\section{More on scaling geometries} \label{sec:scal}

In this section, we discuss further the properties of a scaling geometry with $n \leq 2$ by examining 
the behavior of a probe scalar field. We show that the system has gapless excitations in the IR. We emphasize that here the term IR is used in a relative sense, i.e. IR relative to the UV fixed point. The  understanding of ``genuine'' IR phase of the system depends on how the singularity at $z = \infty$ is resolved. In this sense, the scaling region~\eqref{singf} should be considered as characterizing an intermediate regime, and our discussion of the entanglement entropy of the last section and correlation functions below should be considered as applying only to this intermediate regime. 
In the second part of this section we consider some explicit examples, where a scaling geometry arises as an intermediate phase.

\subsection{Correlation functions}

Consider a probe scalar field in a spacetime~\eqref{1} with~\eqref{singf}. A similar analysis was done  in~\cite{Bianchi:2001de} for two specific flows in $d=4$ dimensions with $n=3$ and $n=2$ respectively,\footnote{There the scalar fields of interest mixed with the metric, here we assume no mixing.} and more recently in~\cite{Dong:2012se} in the context of hyperscaling violating geometries.

The  field equation for a minimally coupled scalar in momentum space can be written as
\be \label{scalarEOM}
 \phi''(z)+\le( { f'(z) \ov 2 f(z)}-{d-1\ov z}\ri) \,\phi'(z)-{m^2 + k^2 z^2\ov z^2 f(z)}\, \phi(z) = 0  \ ,
\ee
where $k^\mu$ is the energy-momentum along the boundary spacetime directions and $k^2 = \eta_{\mu \nu} k^\mu k^\nu$. 

First, consider the gapped case, corresponding to $n>2$. For $z\to\infty$ the two allowed behaviors for the scalar field are:
\bea
\phi_+&=&1-{2k^2\ov (n-2)(2d+n-4)}\, z^{-(n-2)}+\cdots\\
\phi_-&=&z^{d-n/2}-{ 2 k^2\ov (4 + 2 d - 3 n) (-2 + n)}\,  z^{d -n/2- (n-2)} +\cdots \ ,
\eea
where we have set $a=1$ for simplicity of notation.  The null energy condition requires that $n/2<d$~\cite{Dong:2012se}, hence only $\phi_+$ is regular. Near $z \to 0$, the normalizable solution $\phi_{\rm norm} (z)$ can be written as a linear superposition of $\phi_\pm$, i.e. $\phi_{\rm norm} (z) = A_+ (k) \phi_+ + A_- (k) \phi_-$ where $A_\pm (k)$ are some functions of $k^2$. Requiring both regularity at $z \to \infty$ and normalizability at the boundary then 
leads to $A_- (k) = 0$, which implies that the system has a discrete spectrum. 
This is in agreement with the findings of~\cite{Bianchi:2001de} in specific examples, and is consistent with our discussion at the end of last section that such a geometry should be describe a gapped theory.

For $n=2$, in the scaling region~\eqref{scalarEOM} can be solved analytically
\be \label{jpe}
\phi_\pm=\le({m\ov z}\ri)^{-(d-1)/2}\, I_{\pm\nu}\le( {m\ov z}\ri)   
 \qquad \nu = \sqrt{\le({d-1\ov 2}\ri)^2+k^2}\ ,
\ee
where $I$ is the modified Bessel function of the first kind.  For $k^2 < - \De^2$, $\nu$ is imaginary and $\phi_\pm$ behave as plane waves near $z \to \infty$. Then following the standard story~\cite{Son:2002sd}, choosing an infalling solution leads to a complex retarded Green function and a nonzero spectral function. We thus conclude that in this case, there is nonzero gap $\De = {d-1 \ov 2}$ and the system has a continuous spectrum above the gap. 
The presence of a continuum above a gap is presumably responsible for the exponential behavior~\eqref{expsup}
in the entanglement entropy. 


Now we consider $n<2$. For $k^2 < 0$ and $z \to \infty$, the solutions to~\eqref{scalarEOM} have the ``plane wave'' form 
\be
\phi_\pm\to z^{(d-1)/2} \exp\le[\mp i {2 \sqrt{-k^2} \ov 2-n}\,  z^{(2 - n)/2}\ri]  \ .
\ee
Thus in this case one finds a continuous spectrum all the way to $k^2 \to 0_-$. The corresponding spectral function can be extracted from~\cite{Dong:2012se}
\be \label{spectral}
\rho(k^2)=\Im G_R(k^2)\propto (\sqrt{-k^2})^{\ga}, \qquad \ga = {2d-n \ov 2-n} \ .
\ee
This continuous spectrum should be the origin of the ``non-analytic'' behavior in~\eqref{ss1} for a sphere 
and~\eqref{stripan} for a strip. It is also interesting to note that the exponents $\beta$ in~\eqref{stripan},~\eqref{ss1} and $\ga$ in~\eqref{spectral} satisfy a simple relation 
\be 
\ga = \beta + d  \ .
\ee
It would be interesting to understand further the origin of such a relation.

\subsection{Explicit examples: near horizon D$p$-brane geometries}


We now consider the near-horizon Dp-brane geometries~\cite{Itzhaki:1998dd}, which exhibit the scaling geometry~\eqref{singf} in some intermediate regime. EE in these geometries was analyzed previously in~\cite{Dong:2012se}. While these geometries are not asymptotically AdS, 
our earlier result for the non-analytic term in~\eqref{ss1} is nevertheless valid, since it only relies on 
the geometry of the scaling region. We will focus on this leading non-analytic contribution in $1/R$.

 The near horizon extremal black $p$-brane metric in the string frame can be written as 
\bea
ds^2_{string,10}&=&{1\ov \sqrt{g N}}\, \le({r\ov l_s }\ri)^{(7-p)/2}\,(dx^\mu)^2+ \sqrt{g N}\, \le({r\ov l_s}\ri)^{-(7-p)/2}\,\le[dr^2+r^2 d\Om_{8-p}^2\ri]\\
e^{\phi_{10}}&=&g\, (g N)^{-(p-3)/4}\le({r\ov l_s}\ri)^{(p-3)(7-p)/4} \ ,
\eea
where $g$ and $l_s$ are the string coupling and string length respectively. As we will only be interested in the qualitative dependence on $R$ and couplings, here and below we omit all numerical factors. We will restrict our discussion to $p \leq 5$, for which a field theory dual exists. After dimensional reduction and going to the Einstein frame, the metric can be written as (see also~\cite{Dong:2012se})
\be \label{dpb}
ds^2_{Einstein,p+2}={(g N)^{1 \ov p}\,l_s^2\ov z^2}\,\le[(dx^\mu)^2+g N\, {dz^2\ov (z/l_s)^{2(p-3)^2/(9-p)}}\ri] \ ,
\ee
which is of the same form as~\eqref{1} and~\eqref{singf} with
\be \label{veip}
n={2(p-3)^2\ov(9-p)} = \bca  
 1 & p=1 \cr
 {2 \ov 7} & p=2 \cr
 0  & p=3 \cr
 {2 \ov 5} & p=4 \cr
   2 & p=5 
   \eca 
\ , \qquad a = {1 \ov gN l_s^n} \ .
\ee
In our convention, the bulk Newton constant is $G_N= g^2 \, l_s^{d-1}= N^{-2}  (g N)^2\, l_s^{d-1}$.  
The metric~\eqref{dpb} is valid in the range~\cite{Itzhaki:1998dd}
\be \label{imran}
\le(g N\ri)^{-(9-p)/2p}  \ll \le({z\ov l_s} \ri)^{3 - p} \ll  \le(g N^{3-p \ov 7-p}\ri)^{- (9-p)/2p} \ .
\ee
The LHS condition comes from the requirement of small curvature, while the RHS imposes small sting coupling (dilaton). For $p=1,2$ as $z$ is increased 
the system eventually settles into a CFT with degrees of freedom of order $O(N)$ and $O(N^{3 \ov 2})$ respectively, while for $p=4,5$ 
the system is eventually described by the free $U(N)$ Yang-Mills theory (i.e. with $O(N^2)$ degrees of freedom)
as $z \to \infty$. 

For our analysis of the previous section to be valid,  $z_t$ should lie inside the region~\eqref{imran}. For both 
strip~\eqref{rzte} and sphere~\eqref{894} we have $z_t \sim \le(\sqrt{a}\, R\ri)^{2 \ov 2-n}$ which then leads 
to  
\be \label{poep}
{1\ov gN} \ll \le({R \ov l_s} \ri)^{3-p} \ll  {N^{2(5-p)\ov 7-p}\ov gN} \ .
\ee
Now plugging~\eqref{veip} into the ``non-analytic'' term in~\eqref{ss1} for sphere (or similarly~\eqref{stripan}
for strip) we find that 
\be 
\sS \propto N^2 \lam_{eff}^{p-3 \ov 5-p} (R)
\ee
where $\lam_{eff} (R)$ is the effective dimensionless t' Hooft coupling at scale $R$, 
\be 
\lam_{eff} (R) = g N  \le({R \ov l_s}\ri)^{3-p} \ .
\ee
In terms of $\lam_{eff}$ equation~\eqref{poep} can also be written as 
\be 
1 \ll \lam_{eff} (R) \ll N^{2 (5-p) \ov 7-p}  \ .
\ee
For $p=1,2$, $\lam_{eff}$ increases with $R$ but appears in $\sS$ with a negative power. For $p=4$, 
the opposite  happens. In all cases $\sS$ decreases with $R$. 
The $p=5$ case, for which $n=2$, has to be treated differently and one finds from~\eqref{expsup}
\be 
\sS \propto  {N^2\ov (gN)^{3/2}\, \lam_{eff}(R)^{9/2}}\, \exp\le(- {25 \ov8 \lam_{eff} (R)}\ri) \ .
\ee

\section{Domain wall geometry} \label{sec:domain}

We now consider the large $R$ behavior of the REE for holographic systems,
whose IR geometry is described by~\eqref{pep1}, i.e. the system flows to a conformal IR fixed point. 
We will again consider the strip story first. 

\subsection{Strip}

Again we start with~\eqref{z0R} which can be written as 
\be  \label{inr1}
R (z_t) = {z_t\ov \sqrt {f_\infty}}\le[a_d+ \int_{0}^{1} dv \ {v^{d-1} \ov \sqrt{(1-v^{2(d-1)})}}\le({ \sqrt{{ f_\infty \ov f(z_t v)} }}-1\ri)\ri]
\ee
with 
\be 
a_d = {\sqrt{\pi} \Ga \le({d \ov 2 (d-1)} \ri) \ov \Ga \le({1 \ov 2 (d-1)} \ri)} \ .
\ee
The leading behavior in large $z_t$ limit of the integral in~\eqref{inr1} depends on the value of 
\be
\tilde \al \equiv \tilde \De - d \ .
\ee
For $\tilde \al < {d \ov 2}$ we can directly expand $f(z_t v)$ using~\eqref{find1} 
\be \label{onre}
\sqrt{{ f_\infty \ov f(z_t v)} }-1 = {1 \ov 2 f_\infty (\tilde \mu z_t v)^{2 \tilde \al}} + \cdots 
\ee
and find 
\be \label{rzt1}
R (z_t) = {z_t\ov \sqrt {f_\infty}}\le[a_d+ {\tilde b_d \ov (\tilde \mu z_t)^{2\tilde \al}} + \cdots \ri] \ 
\ee
with 
\be 
\tilde b_d = {\sqrt \pi \,\Gamma\le(d-2\tilde\al \ov 2(d-1)\ri) \ov 2 f_\infty (1-2\tilde\al) \,\Gamma\le(1-2\tilde\al \ov 2(d-1)\ri)}\ .
\ee
Note that $\tilde b_d$ is positive for any $d > 1$. 
For $\tilde \al \geq {d \ov 2}$, the term on RHS of~\eqref{onre} leads to a divergence in~\eqref{inr1} near $v=0$
and should be treated differently.\footnote{Note that even for $\tilde \al < {d \ov 2}$, higher order terms in the expansion on the RHS of~\eqref{onre} can similarly lead to divergences. They can be treated similarly as 
for $\tilde \al \geq {d \ov 2}$, and give rise to higher order terms compared to the second term of~\eqref{rzt1}.} In particular,  the divergence indicates that the leading contribution should 
come from the integration region $v \ll 1$. We will thus approximate the factor $1/\sqrt{1- v^{2 (d-1)}}$ in the integrand of~\eqref{inr1} by $1$, leading to  
\be \label{rzt2}
R (z_t) = {z_t\ov \sqrt {f_\infty}}\le(a_d+ {b_d \ov  z_t^{d} } + \cdots\ri) 
\ee
where 
\be \label{coebd}
b_d = \int_{0}^{\infty} du \ u^{d-1} \le({ \sqrt{{ f_\infty \ov f(u)} }}-1\ri) \ . 
\ee

Inverting~\eqref{rzt1} and~\eqref{rzt2} we find from \eqref{s2} 
\be \label{strid}
\sR_d={L^{d-1} \ov 2 G_N} \le({a_d \ov \sqrt{f_\infty}}\ri)^{d-1} \times
\begin{cases}
1 + (d-1) {\tilde b_d \ov a_d} \le({\tilde \mu \sqrt{f_\infty} R \ov a_d}\ri)^{- 2 \tilde \al} + \cdots & 
(\tilde \al<{d\ov2}) \cr
1 + (d-1) { b_d \ov a_d} \le({\sqrt{f_\infty} R \ov a_d}\ri)^{-d} + \cdots & 
(\tilde \al \geq {d\ov2}) 
\eca \ . 
\ee
We discuss the physical implication of this result in Sec.~\ref{sec:diss}.

\subsection{Sphere}

With~\eqref{pep1} as $z \to \infty$ the system flows to a CFT in the IR, and, as discussed in~\cite{Liu:2012eea},
to leading order in the large $R$ expansion the REE $\sS_d$ approaches a constant, that of the IR CFT. 
Here we confirm that the subleading terms have the structure given in~\eqref{irex}.  

\subsubsection{IR expansion}

Since the IR geometries approaches  AdS, in the large $R$ limit the IR part of the minimal surface should 
approach that in pure AdS. In particular, in the limit $R \to \infty$, we expect most part of the minimal surface 
to lie in the IR AdS region, hence the IR solution $z (\rho)$ can be written as  
\be \label{inew}
z(\rho)=z_0(\rho)+ z_1 (\rho) + \cdots , \qquad z_0(\rho)=\sqrt{f_\infty\le(R^2- \rho^2\ri) } \ .
\ee
$z_0 (\rho)$ is the minimal surface with boundary radius $R$  in a pure AdS with $f = f_\infty$. $z_1$ and $\cdots$ in~\eqref{inew} denote subleading corrections which are suppressed compared with $z_0$ by some inverse powers of $R$. Below we will determine the leading correction $z_1 (\rho)$ by matching with the UV solution. 


Plugging~\eqref{inew} into~\eqref{rhoeom}, and expanding to linear order in $z_1$, we find that
\be
 z_1''+{(d-2)R^2-2 \rho^2 \ov \rho(R^2-\rho^2)}z_1'-{(d-1)R^2 \ov (R^2-\rho^2)^2}z_1= s(\rho) \ ,  \label{perteq} 
\ee
where the source term $s(\rho)$ is given by 
\be
s(\rho)={f_\infty^{\ha-\tilde\al}\ov \tilde\mu^{2\tilde\al}}\, {(d-1)R^2 + (\tilde\alpha -1) \rho^2 \ov ({R^2- \rho^2})^{3/2+\tilde\alpha}}  \ .
\ee
The homogenous equation, obtained by setting $s (\rho)$ to zero in~\eqref{perteq}, 
has the following linearly independent solutions 
\bea
\phi_1&=& {R \ov \sqrt{R^2-\rho^2}} \\
\phi_2&=& \begin{cases}
{3\ov2}\le[-1+ {R \ov \sqrt{R^2-\rho^2}} \ {\rm arctanh}\le({ \sqrt{R^2-\rho^2} \ov R}\ri)\ri] \qquad\qquad\qquad (d=3) \\
{(R-\rho)^2\ov \sqrt{2}\,\rho  \sqrt{R^2-\rho^2}}  \qquad\qquad\qquad\qquad\qquad\qquad\qquad\qquad (d=4) \\
{5(R^2+2\rho^2) \ov 8\rho^2} -\frac{15}{8} \,{R \ov \sqrt{R^2-\rho^2}} \ {\rm arctanh}\le({ \sqrt{R^2-\rho^2} \ov R}\ri) \qquad\quad (d=5) \\
\end{cases}\\
W(\phi_1,\phi_2)&\equiv &\phi_1 \phi_2'-\phi_1' \phi_2= \begin{cases} 
-{3R \ov 2\rho \sqrt{R^2-\rho^2}} \qquad (d=3)\\
-{R\ov\sqrt{2}\, \rho^2} \qquad\qquad\quad (d=4)\\
-{5R\sqrt{R^2-\rho^2}\ov 4\rho^3} \qquad (d=5)
\end{cases}  \ .
\eea
Note that there is an expression for $\phi_2$ in terms of hypergeometric functions for all dimensions, but 
we find it more instructive to display explicit expressions in various dimensions. 
The final results will be written down in general $d$.  $\phi_1$ is singular at $\rho = R$, while $\phi_2 \sim {R^{d-3} \ov \rho^{d-3}}$  is singular as $\rho \to 0$ (for $d=3$, there is a logarithmic divergence) with 
$W \to - {R^{d-3} \ov \rho^{d-2}} + \cdots$. Also note that 
\be
\phi_2 \to \de^{d-1 \ov 2}, \qquad W\to -{d\ov \sqrt{8}} \,{\de^{d-4 \ov 2} \ov R}  ,  \qquad
\de \equiv {R- \rho \ov R} \ll 1 \ .
\ee

 In order for $z(\rho)$ to be regular at $\rho=0$, 
$z_1$ should be regular there, and can be written as  
\be \label{onew}
\begin{split}
z_1(\rho)&=
c R \phi_1(\rho)+ \phi_1(\rho) \int_\rho^{R} dr \ { \phi_2(r) \ov W(r)}s(r)+\phi_2(\rho) \int_0^\rho dr \ {\phi_1(r)  \ov W(r)}s(r)   \ ,
\end{split}
\ee
where $c$ is an integration constant. Note that the first integral above is convergent in the upper integration limit only for $\tilde \alpha < 1$. For $\tilde \al \geq 1$ some additional manipulations are required. For example 
for $1<\tilde\al<2$, we should replace the first integral by
\be
\begin{split}
& \phi_1(\rho) \, \le[\int_\rho^{R} dr \ \le( {\phi_2(r) \ov W(r)}s(r) - 
{c_1 \ov \tilde \mu^{2 \tilde \al} R^{\tilde\al} (R-r)^{\tilde\al} }\, \ri)
+{c_1 R \ov (\tilde\al-1)(\tilde \mu R)^{2 \tilde\al}}\, {1\ov \de^{\tilde\al-1}}\ri] 
\end{split}
\ee
where $c_1$ is the numerical constant appearing in the limit $ {\phi_2(r) \ov W(r)}s(r)
\to {c_1 \ov \tilde \mu^{2 \tilde \al} R^{\tilde \al} (R-r)^{\tilde \al}} + \cdots$ as $r \to R$,
and is given by 
\be
c_1=-{2^{-\tilde\al} f_\infty^{1/2 - \tilde \al} (-2 + d + \tilde \al)\ov d} \ .
\ee
 For $\tilde\al>2$  further subtractions may be needed. We will not write these separately, as they are irrelevant for our discussion below.

\subsubsection{Matching}

The IR expansion~\eqref{inew} and~\eqref{onew} is valid for $z \gg z_{CO}$, where~\eqref{find1} applies. 
For sufficiently large $R$, this includes the region where
\be \label{mheo}
\de \equiv {R- \rho \ov R} \ll 1, \qquad R \sqrt{\de} \gg z_{CO}, \tilde \mu^{-1}, \cdots  \
\ee
where the $\cdots$ on the right hand side of the second inequality includes all other scales of the system. 
The UV expansion we discussed earlier in Sec.~\ref{sec:uvexp} applies to the region 
$\de \ll 1$. Thus the IR and UV expansions can be matched for $\rho$ satisfying~\eqref{mheo}.

Let us now consider the behavior of~\eqref{onew} in the overlapping region~\eqref{mheo}. The first integral 
gives 
\be
\phi_1(\rho) \int_\rho^{R} dr \ { \phi_2(r) \ov W(r)}s(r) 
= d_1 R  {\de^{1/2-\tilde\al} \ov (\tilde \mu R)^{2 \tilde \al}} \le(1+ O(\de) \ri) \ ,
\ee
where for all $\tilde \al$ 
\be 
d_1 = -{(d-2+\tilde\al) \le(2 f_\infty \ri)^{1/2-\tilde\al} \ov 2(1-\tilde\al) d}  \ .
 \ee
 The second integral in~\eqref{onew} gives 
\be
\phi_2(\rho) \int_0^\rho dr \ {\phi_1(r)  \ov W(r)}s(r) =
d_2 R {\de^{1/2-\tilde\al} \ov (\tilde \mu R)^{2 \tilde \al}} \le(1+ O(\de) \ri)
+ h R {\de^{d-1 \ov 2} \ov (\tilde \mu R)^{2 \tilde \al}} \le(1+ O(\de) \ri) \ ,
\ee
where
\be 
d_2 = -{d-2+\tilde\al\ov d(d-2+2\tilde\al)} \, \le(2 f_\infty \ri)^{1/2-\tilde\al}, \quad
h =  f_\infty^{1/2-\tilde\al} {2^{(d-3)/2}\pi\tilde\al \, \Gamma\le({d+1\ov 2}\ri)\ov d\sin\le({\pi\ov 2}(d+2\tilde\al) \ri)\Gamma\le(\frac32-\tilde\al\ri)\Gamma\le(\frac d2+\tilde\al\ri)}\ . 
\ee
Putting the two expansions together we get:
\be
 \label{hdef}
z_1(\rho)={c R\ov \sqrt{2\delta}} + d_3 R {\de^{1/2-\tilde\al} \ov (\tilde \mu R)^{2 \tilde \al}} \le(1+ O(\de) \ri)
+ h R {\de^{d-1 \ov 2} \ov (\tilde \mu R)^{2 \tilde \al}} \le(1+ O(\de) \ri) \ ,
\ee
where 
\be 
d_3 =
-{d-2+\tilde\al\ov 2(1-\tilde\al)(d-2+2\tilde\al)} \, \le(2 f_\infty \ri)^{1/2-\tilde\al} \ .
\ee

One could consider the next order in the IR expansion, i.e. including a $z_2$ in~\eqref{inew}. The equation for $z_2$ only differs from~\eqref{perteq} by having a different source term, and
the corresponding terms in~\eqref{hdef} coming from the source will be proportional to $(\tilde \mu R)^{- 4 \tilde \al}$. Similarly,  the corresponding terms at the $n$th order are proportional $\le(\tilde \mu R \ri)^{-2 n \tilde\al}$.

Now including $z_0$ in the region~\eqref{mheo}, we have the expansion 
\be
 \label{irexp}
{z (\rho) \ov R} =\sqrt{2 f_\infty\, \delta}\le[1+{c\ov \sqrt{4 f_\infty}\, \delta} + {d_3\ov \sqrt{2 f_\infty}}  {1  \ov (\tilde \mu R\,  \sqrt{\delta})^{2\tilde\al}}+ 
{h\ov \sqrt{2 f_\infty}}\, {\delta^{(d+2\tilde\al-2)/2}\ov (\tilde \mu R \,  \sqrt{\delta})^{2\tilde\al}}+\cdots\ri] \ .
\ee
Clearly we have a double expansion in terms of $\de$ and inverse powers of $\tilde \mu R \,  \sqrt{\delta}$.
The consistency of the expansion also requires that the constant $c$ have the scaling 
\be 
c = {\tilde c \ov (\tilde \mu R)^2}  
\ee
with $\tilde c$ now an $O(R^0)$ constant. 
Now inverting~\eqref{irexp} we find that 
\be \label{diexp}
\de = {z^2 \ov 2f_\infty R^2} - {\tilde c \ov \sqrt{f_{\infty}} (\tilde \mu R)^2} + {d_4 \ov 2 f_\infty} 
 {z^2 \ov R^2} {1 \ov (\tilde \mu z)^{2\tilde\al}}- 
\tilde h {z^d \ov \tilde \mu^{2 \tilde \al} R^{d + 2 \tilde \al}} \,+ \cdots \ ,
\ee 
which can be considered as a double expansion in ${z / R}$ and $1/(\tilde \mu z)$ and 
\be 
d_4 = 
 {d-2+\tilde\al\ov (1-\tilde\al)(d-2+2\tilde\al)} \ , \qquad \tilde h = 2h (2 f_\infty)^{-{d+1 \ov 2}} \ .
\ee

Now consider~\eqref{rho1s} with $z$ large, with $f (z)$ given by~\eqref{find1}.
We find that $\rho_1$ can be expanded as  (see Appendix~\ref{sec:exp} for details)
\be
\label{uvexp1}
\rho_1(z)=
{b_1 \ov d \sqrt{f_\infty}} z^d\le(1
+\cdots\ri) + {z^2\ov 2 f_\infty}+\gamma + {z^2\ov 2 f_\infty} a 
(\tilde \mu z)^{-2\tilde\al} + O\le({z^2 \ov (\tilde \mu z)^{4 \tilde \al}} \ri)  \ . 
\ee
Note the above equation applies to all $\tilde \al$, but the expression for constant $\ga$ depends on the 
range of $\tilde \al$. For example, for $\tilde \al > 1$, 
\be 
\ga = \int_{0}^\infty du \ \le[(d-2) {u^{d-1}\ov \sqrt{f(u)}} \int_{u}^\infty dv \ {1 \ov v^{d-1}\sqrt{f(v)}}-{u\ov f_\infty}\ri] 
\ .
\ee
At higher orders in $1/R$, it suffices to determine the leading term:
\be \label{opw}
\rho_n(z)={ b_n \ov d \sqrt{f_\infty}} z^d + \cdots  \qquad\qquad \hat\rho(z)={\hat b \ov d \sqrt{f_\infty}} z^d + \cdots \ .
\ee
Using~\eqref{uvexp1} and~\eqref{opw} in~\eqref{exp} we find that 
\be 
\de =  {b_1 \ov d \sqrt{f_\infty} R^2} z^d  (1 + \cdots) + {z^2 \ov 2 f_\infty R^2} \le(1 + a (\tilde \mu z)^{- 2 \tilde \al} + \cdots \ri) + {\ga \ov R^2} + \cdots  + {\hat b \ov d \sqrt{f_\infty}} {z^d \ov R^{\nu}} + \cdots \ .
\label{alexp}
\ee
Comparing~\eqref{alexp} with~\eqref{diexp} we find they match provided that 
\be \label{core0}
 b_1 = 0, \qquad \tilde c = - \sqrt{f_\infty} \tilde \mu^2 \ga , \qquad \hat b = - d \sqrt{f_\infty} {\tilde h \ov \tilde \mu^{2 \tilde \al}}, \qquad
\nu = d + 2 \tilde \al -1  \ .
\ee

\subsubsection{Asymptotic expansion of REE}

With $\hat b$ and $\nu$ given by~\eqref{core0}, from~\eqref{eooe} we find 
the leading ``non-analytic'' contribution in $\sS_d$ is given by 
\be
\sS_d= \cdots +  
\ha K_{IR} {(d-1)!!\ov (d-2)!!}\,  b(\tilde\al) \, { f_\infty^{-\tilde\al} \ov (\tilde \mu R)^{2\tilde\al}}+\cdots 
\ee
with 
\be \label{bsign}
K_{IR} \equiv K f_\infty^{-(d-1)/2} , \qquad  b(\tilde\al)=\begin{cases}
 {1\ov 1-2\tilde\al} 
 & \text{$d$ odd}\cr
 {\sqrt{\pi}\, \Gamma\le(1-\tilde\al\ri)\ov 2 \Gamma\le(\frac32-\tilde\al\ri)} 
&  \text{$d$ even} 
 \end{cases} \ .
\ee
This above expression agrees with that obtained in~\cite{Liu:2012eea} for two closely separated fixed points, which we review and extend in Appendix~\ref{sec:close}. As discussed in the Introduction this can be anticipated on the grounds that the coefficient of the non-analytic term should depend only on the physics at the IR fixed point. 

As discussed earlier our UV expansion~\eqref{exp} was designed to produce the second line of~\eqref{irex}, 
and the fact that the UV expansion is consistent with the IR expansion confirms the second line of~\eqref{irex}.

In $d=3$ using $\rho_1$ and $z_1$ obtained in last subsection we can
obtain the coefficient of $1/R$ term by directly evaluating the action as we have done for the gapped and scaling geometries.  The calculation is given in Appendix~\ref{dwonshell}. The final answer is:
\be\label{s3Ftheorem}
\sS_3=\sS_3^{(IR)}+K_{IR} \, { f_\infty^{-\tilde\al} \ov (1-2\tilde\al)(\tilde \mu R)^{2\tilde\al}}+{Ks_{1}\ov \tilde \mu R}+\dots \ , 
\ee
where $s_{1}$ is given by~\eqref{dwresult}:
\bea
s_{1}&=& \begin{cases}
\int_0^{\infty} dz \ \le[{z^2 \ov \sqrt{f(z/\tilde \mu)}} \le[\int_z^\infty dv \ {1\ov v^2 \sqrt{f(v/\tilde \mu)}}\ri]^2-{1 \ov f_\infty^{3/2}} \ri] \qquad\qquad\qquad \le(\ha<\tilde\al\ri) \\
\int_0^{\infty} dz \ \le[{z^2 \ov \sqrt{f(z/\tilde \mu)}} \le[\int_z^\infty dv \ {1\ov v^2 \sqrt{f(v/\tilde \mu)}}\ri]^2-{1 \ov f_\infty^{3/2}}\le(1+{3+2\tilde\al\ov 2(1+2\tilde\al)}\, {1 \ov z^{2\tilde\al}}\ri) \ri] \qquad\qquad \le(\frac14<\tilde\al<\ha\ri)
\end{cases} \ . \label{dwresultsmain}
\eea
The expressions for smaller values of $\tilde\al$ are similar but require more subtractions. $s_1$ (and the integration variables, $z$ and $v$) is dimensionless, hence only depends on ratios of RG scales. 

Our results are compatible with the F-theorem; for $\tilde\al<\ha$ the non-analytic term dominates in~\eqref{s3Ftheorem}, and $b(\tilde \al)>0$ in this range~\eqref{bsign}. For $\ha<\tilde\al$, where the $1/R$ term dominates over the non-analytic term, $s_1>0$ follows from~\eqref{dwresultsmain}.

 As a consistency check, 
we apply these formulae to closely separated fixed points in  Appendix~\ref{sec:close}. We recover~\eqref{closely} that is obtained using different methods. Another consistency check is that the $f_\infty\to\infty$ limit of~\eqref{dwresultsmain} recovers $s_1$ for the scaling geometries~\eqref{a12}. This had to be the case, as a scaling geometry can be viewed as a limit of domain walls with increasing $f_\infty$.

\subsection{Discussion} \label{sec:diss}

We conclude this section making a comparison between the result for the strip~\eqref{stipREE},~\eqref{strid} and 
that for the sphere~\eqref{irex}.

First, let us look at the strip result~\eqref{strid}. When $\tilde \al < {d \ov 2}$, $\sR_d$ can be written in terms of 
an effective dimensionless irrelevant coupling $g_{eff} (R) = (\tilde \mu R)^{-\tilde \al}$ as 
\be 
\sR_d = \sR_d^{(IR)} + \# g_{eff}^2 (R)  + \cdots
\ee
with a coefficient $\#$ only depending on the data at the IR fixed point. 
As for the sphere case~\eqref{irex1}, such a term can be expected from conformal perturbations around a fixed point.  For $\tilde \al > {d \ov 2}$, we see that the leading approach to the IR value saturates at $R^{-d}$ no matter what the dimension of the leading irrelevant operator is. In particular, the coefficient $b_d$~\eqref{coebd} involves an integral over all spacetime, suggesting this term receives contributions from degrees of  freedom at all length scales (not merely IR degrees of freedom). This term may be considered as the counterpart for a strip of the second line in~\eqref{irex}. But note that for a sphere the second line of~\eqref{irex} can be associated with a curvature expansion of a spherical entangling surface, while for a strip all such curvature terms are absent.

\section{Black holes}\label{sec:bh}

In this section we consider the large $R$ expansion of the entanglement entropy for strip and sphere 
for a holographic system at a finite temperature/chemical potential, which is described 
by a black hole on the gravity side. Compared with examples of earlier sections, there are some new elements in the UV and IR expansions. The setup is exactly the same as discussed in Sec.~\ref{sec:eest} and Sec.~\ref{sec:eesp} except that now the function $f (z)$ has a zero at some $z = z_h$: 
\be 
f (z_h) =0 , \qquad f (z) = f_1 (z_h - z) + f_2 (z- z_h)^2 + \cdots , \quad z \to z_h \ . 
\ee
In our discussion below, we will assume $f_1$ is nonzero. 
For an extremal black hole, $f_1$ vanishes, which requires a separate treatment and will be given elsewhere.
For notational simplicity, we will set $z_h =1$ below, which can be easily reinstated on dimensional grounds.
We also introduce 
\be
 \label{defga}
\ga \equiv \sqrt{(d-1) f_1 z_h \ov 2}\ , \qquad
\ee
which will appear in many places below. 

\subsection{Strip}

We again look at the strip first. As $R \to \infty$ we expect the tip of the minimal surface $z_t$ to 
approach the horizon $z_h=1$. This can be seen immediately from equation~\eqref{z0R}: 
with $z_t = 1$, due to $f (1)=0$, the integrand develops a double pole and the integral becomes
divergent. To obtain the large $R$ behavior, we thus take 
\be
z_t =1 - \ep, \qquad \ep \ll 1 \ ,
\ee
and expand the integral in $\ep$. 
From~\eqref{z0R} we find that 
\be 
{R }= - {1 \ov 2 \ga} \log {\ep \ov 4} + b_0 + O(\ep \log \ep)  \ ,
\ee
where $\ga$ was introduced in~\eqref{defga} and 
\be \label{defb0}
 b_0 = 
  \int_{0}^{1} dv \ \le[{v^{d-1} \ov \sqrt{f\le(v\ri) (1-v^{2(d-1)})}}-{1\ov 2\gamma}\,  {1 \ov \le(1- v\ri)}\ri] \ .
\ee
Then we can express $\ep$ as a function of $R$:
\be 
\ep = 4 e^{2 \ga b_0} e^{-  {2 \ga R }} \le(1 + O(R e^{-2 \ga R}) \ri)  \ .
\ee

Reinstating $z_h$, from~\eqref{s2}
\be 
\sR_d = {L^{d-1} \ov 2 G_N} \le({R \ov z_h}\ri)^{d-1}  \le(1 + (d-1) \ep + \cdots \ri) \ .
\ee
The entanglement entropy itself can be written as 
\be 
S_{\rm strip} =  {L^{d-1}  \ov 4 G_N} \, {2R \, l^{d-2}\ov z_h^{d-1}} \le(1 - { 2(d-1) z_h \ov  \ga R} e^{2 \ga b_0} e^{-  {2 \ga R \ov z_h}}  + \cdots \ri)  \ ,
\ee
which is given by the Bekenstein-Hawking entropy with exponential corrections. 
For the $d=2$ BTZ black hole one simply recovers the well known expression for a 2d thermal 
CFT by evalutaing~\eqref{z0R} exactly.

\subsection{Sphere}

\subsubsection{UV expansion}\label{sec:uvexpBH}

Anticipating a volume term and possibly other subleading terms in the entanglement entropy, we modify the 
UV expansion~\eqref{exp} to include terms of all integer powers in $1/R$, i.e. 
\be \label{Rexp0}
\rho (z) = R - \rho_0 (z) - {\rho_1 (z) \ov R} + \cdots \ .
\ee
At finite temperature, we do not expect non-integer power law terms in $1/R$ in~\eqref{Rexp0}, except exponentially small terms. Here will focus on the lowest two terms in~\eqref{Rexp0}.

The equations for $\rho_0$ and $\rho_1$ are 
\bea 
&& \rho_0'' + {f' \ov 2f} \rho_0' - {d-1 \ov z} \rho_0' (1 + f \rho_0'^2) = 0
\cr
&& \rho_1'' + \le({f ' \ov 2f} - {(d-1) (1+3 f \rho_0'^2) \ov z} \ri) \rho_1' 
+ {d-2 \ov f} (1 + f  \rho_0'^2)=0 \ ,
\eea
which can be solved by 
\be \label{sol1}
\rho_0 = \int_0^z dy \, {y^{d-1} \ov f^\ha \sqrt{a^{-1} - y^{2 (d-1)}}}
\ee
and 
\be \label{sol2}
\rho_1  (z) =\int_0^z dz \, {z^{d-1} \ov f^\ha \le(1- a z^{2 (d-1)}\ri)^{3 \ov 2}} \le(b + (d-2) \int_z^{1} dy \, {\sqrt{1- a y^{2 (d-1)}} \ov f^\ha y^{d-1}} \ri)
\ee
with $a$ and $b$ integration constants. 

The expansion~\eqref{Rexp0} should break down for small $\rho$ when $\rho_0$ or higher order terms become comparable to $R$. As in the strip case we again expect that the tip of the surface $z (\rho=0) \equiv z_t$ approaches the horizon $z=1$, when $R$ is large. We thus expect the UV expansion to break down near the horizon. This indicates that we should choose 
\be \label{onne}
a = 1 \ . 
\ee
An immediate consequence of the above equation is that the expansion of $\rho_0$ near the boundary has the form 
\be 
\rho_0 = {1 \ov d} z^d +\cdots  \quad \to \quad 
c_d(R)=-{1\ov d}+\cdots \ , \label{bhc}
\ee
which from~\eqref{eopr} immediately gives 
\be 
S = {L^{d-1} \ov 4 G_N} {\om_{d-2} \ov d-1} {R^{d-1} \ov z_h^{d-1}} + \cdots
=  {L^{d-1} \ov 4 G_N} {V_{\rm sphere}  \ov z_h^{d-1} }+ \cdots \ ,
\ee
where $V_{\rm sphere}$ is the volume of the sphere and we have reinstated $z_h$. 
In Sec.~\ref{sec:arb} we generalize this result to an arbitrary shape.

\subsubsection{IR expansion}

It is clear both from general arguments and the numerical solution shown in Fig.~\ref{qualitative} that the IR part of the minimal surface is very flat and stays in the near horizon region for a large range of $\rho$. This motivates 
us again to write 
\be
z_t =1 - \ep \qquad \ep \to 0 \ .
\ee
 
The part of minimal surface near the horizon can then be expanded in terms of $\ep$
\be \label{horex}
z (\rho) = 1 - \ep z_1 (\rho) - \ep^2 z_2 (\rho) + \cdots \ 
\ee
with  boundary conditions 
\be 
z_1 (0) = 1, \qquad z_m (0) = 0, \;\; m \geq 2, \qquad z_n' (0) = 0  , \;\; n \geq 1 \ .
\ee
Below we will relate $\ep$ to $R$ by matching~\eqref{horex} with the UV expansion~\eqref{Rexp0}. 

Plugging~\eqref{horex} into the equation of motion~\eqref{eom3} we find that $z_1$ satisfies the equation 
\be 
 {z_1'' \ov z_1} -\ha {z_1'^2 \ov z_1^2} + {d-2 \ov \rho} {z_1' \ov z_1} - {\ga^2 \ov 2 } = 0 \ ,
 \ee 
 where $\ga$ was introduced in~\eqref{defga}. Setting $z_1 = h^2$, one finds that $h$ satisfies the Bessel equation which then leads to
\be 
z_1 =  \Ga^2 \le({d-1 \ov 2} \ri)\le({\ga \rho \ov 2}\ri)^{3-d} I_{d-3 \ov 2}^2 (\ga \rho) \ ,
\ee
where we have imposed the boundary condition at $\rho=0$. At large $\rho$ we then find that 
\be \label{hie0}
z_1 = \Ga^2 \le({d-1 \ov 2} \ri)\le({\ga \rho \ov 2}\ri)^{3-d} {e^{2 \ga \rho} \ov 2 \pi \ga \rho} \le(1 + O(\rho^{-2}) \ri) 
\ . 
\ee

\subsubsection{Matching}

We now try to match the two sets of expansions in their overlapping region with 
\be \label{ovreg}
1 \ll \sig \equiv R - \rho \ll R, \qquad \ep \ll u \equiv 1- z \ll 1 \ .
\ee
In the above region equation~\eqref{hie0} can be expanded in large $R$ as 
\be 
z_1 =  \Lam e^{- 2 \ga \sig} \le(1 + {c_{11} (\sig) \ov R} + {c_{12} (\sig) \ov R^2} + \cdots \ri)
\equiv C_1 (\sig) \Lam e^{- 2 \ga \sig}
\ee
with 
\be 
\Lam = \Ga^2 \le({d-1 \ov 2} \ri){2^{d-4}  \ov \pi \ga^{d-2} } {e^{2 \ga R} \ov R^{d-2}}
\qquad c_{11} (\sig) = (d-2)   \sig, \qquad \cdots  \ .
\ee
One can show that $z_2$ has a similar structure, i.e. 
\begin{align}
z_2 =&   \Lam^2 e^{- 4 \ga \sig} c_{20} \le( 1+ {c_{21} (\sig) \ov R} + {c_{22} (\sig) \ov R^2} + \cdots \ri)
\equiv \Lam^2 e^{- 4 \ga \sig} C_2 (\sig) \ . 
\end{align}
We thus have 
\be \label{eio2}
u = \ep z_1 + \ep^2 z_2 + \cdots  
= \ep \Lam C_1 (\sig) e^{- 2 \ga \sig} + (\ep \Lam)^2 C_2 (\sig) e^{- 4 \ga \sig} + \cdots  \ .
\ee

One now expands $\rho_0$ and $\rho_1$ for small $u$
\bea \label{onex0}
\rho_0 &= & - {1 \ov 2 \ga } \log u + b_{0} + b_{01} u + \cdots  , \\ 
\label{onex1}
\rho_1 &= &  {b \ov 4 \ga (d-1)} {1 \ov u} -b_\text{log} \log u + b_{10} + b_{11} u + \cdots 
\eea
where various coefficients $b_{0}, b_{01}, \cdots$ can be found explicitly from~\eqref{sol1}--\eqref{sol2}.
In particular $b_0$ is given by~\eqref{defb0}. 
Using~\eqref{onex0}--\eqref{onex1} in~\eqref{Rexp0} we then find that  
\be \label{two0}
\sig =  \rho_0 + {\rho_1 \ov R} + \cdots 
=  {b \ov 4 \ga (d-1)} {1 \ov u} - {1 \ov 2 \ga} B_c (R) \log u + B_0 (R) + B_1 (R) u + \cdots  \ ,
\ee
where 
\be 
B_c (R) = 1 +  { 2 \ga b_\text{log} \ov R} + O(R^{-2}) , \qquad B_0 (R) = b_{0} + {b_{10} \ov R} + O(R^{-2}), \qquad \cdots \ .
\ee

Now matching~\eqref{eio2} and~\eqref{two0} we find they precisely match provided that $b =0$ and 
\be 
\ep = \ep_0 \le(1 + {d_1 \ov R} + {d_2 \ov R^2} + \cdots \ri) 
\ee
with 
\be 
\ep_0 = e^{2 \ga b_0} \Lam^{-1} = \le(\Ga^2 \le({d-1 \ov 2} \ri){2^{d-4}  \ov \pi \ga^{d-2} } {e^{2 \ga R} \ov R^{d-2}}\ri)^{-1}
e^{2 \ga b_{0}}, \qquad d_1 = 2 \ga b_{10} - (d-2) b_{0} , \quad \cdots \ .
\ee

\subsection{Large $R$ behavior of the entanglement entropy}

By carrying out the procedure outlined above one could in principle obtain the large $R$ 
expansion for the entanglement entropy to any desired order. 
As an illustration we now calculate the constant term (i.e. $R$-independent term) in $S$ for $d=3$. 

We divide the area functional~\eqref{sphact} into a UV and IR piece and calculate to $O(R^0)$:
\bea
A&\equiv& A_{UV}+A_{IR} \\
A_{UV}&=&\int_\delta^{z_*} dz \ {\rho \ov z^{2}} \sqrt{\rho'(z)^2+{1\ov f(z)}}\\
A_{IR}&=&\int_{0}^{\rho_*} d\rho \ {\rho \ov z^{2}}   \sqrt{1+{z'(\rho)^2\ov f(z)}} \ ,
\eea
where $z_*$ is an arbitrary point in the matching region and $\rho(z_*)=\rho_*$ and $\de$ is a UV cutoff. Plugging in the UV expansion \eqref{Rexp0} and \eqref{sol1} into $A_{UV}$  we get:
\be
A_{UV}=\int_\delta^{z_*} dz \ {R-\rho_0(z)\ov z^2\sqrt{f(z)(1-z^4)}}+\rho_1(z_*)+O\le({1\ov R}\ri)
\ee
This has an expression for small $u_*=1-z_*$:
\bea
A_{UV}&=&-{1\ov 8 \ga^2} \,\log^2 u_* +  {R-b_{0}\ov 2 \ga}\,\log u_* +\rho_1(u_*)+{R\ov \delta}+a_{UV}+O(u_*) \label{AUVbh}\\
a_{UV}&\equiv&-R+\int_0^{1} dz \ \le[ {R-\rho_0(z)\ov z^2\sqrt{f(z)(1-z^4)}} -{R\ov z^2}-{1\ov 4 \ga^2}\, {\log (1-z)\ov (1-z)}-{1\ov 2 \ga}\, {R-b_{00}\ov 1-z}\ri]  \ .
\eea
Note that $\rho_1(u_*)$ contains $\log u_* $ and constant terms, but we chose not to expand it for later convenience.  We isolated all $u_*$ and $\delta$ dependence, hence $a_{UV}$ is a finite term independent of $u_*$. It includes finite area law terms. 
$A_{IR}$ is  given by 
\be \label{AIRexp}
A_{IR}=\int_{0}^{\rho_*} d\rho \ \le[\rho+\ep \, \rho\le(2 z_1(\rho)+{z_1'(\rho)^2\ov 2 \ga^2 z_1(\rho)}\ri)+O(\ep^2)\ri]\ .
\ee
Plugging in the results of the IR expansion we find 
\be
\begin{split} \label{AIRbh}
A_{IR}&={\rho_*^2\ov 2}+O(u_*) = {R^2\ov 2}-R\rho_0(u_*)+{\rho_0(u_*)^2\ov 2}-\rho_1(u_*)+O(u_*)\\
&={R^2\ov 2}+{1\ov 8 \ga^2} \,\log^2 u_* -  {R-b_{0}\ov 2 \ga}\,\log u_*+{b_{0}^2\ov2}-\rho_1(u_*)\ .
\end{split}
\ee
Adding together \eqref{AUVbh} and \eqref{AIRbh}, we find that the $u_*$ dependence cancels which provides a nontrivial  consistency check, and the final result is
\bea
A&=&\#{R^2\ov 2}+\le(\text{area law terms}\ri)+a\\
a& = &{b_{0}^2\ov2}-\int_0^{1} dz \ \le[ {\rho_0(z)\ov z^2\sqrt{f(z)(1-z^4)}} -{1\ov 2 \ga}\, {-{1\ov 2 \ga}\log (1-z)+b_{0}\ov (1-z)}\ri]  \ . 
\eea
$b_0$ is the constant term in the expansion~\eqref{onex0} of $\rho_0$, and it is given by~\eqref{defb0}.

\subsection{Leading order result for an arbitrary shape} \label{sec:arb}

For arbitrary shape we cannot go into as much detail as for the sphere case. Here we demonstrate that at leading order in the large size limit the entanglement entropy goes to thethermal entropy in an explicit calculation. To the best of our knowledge this is the first demonstration using the holographic approach, although the result is widely expected. 

We choose spherical coordinates on each $z$ slice of the spacetime:
\bea
ds^2\vert_{t=0}&=&{L^2\ov z^2}\le(\le[d\rho^2+\rho^2d\Omega^2_{d-2}\ri]+{dz^2\ov f(z)}\ri)\\
d\Omega^2_{d-2}&=&\sum_{i=1}^{d-2} g_i \, d\theta_i^2 \ ,
\eea
where $g_i$ are just the conventional metric components:
\be
g_1=1\ , \qquad g_2=\sin^2\theta_1\ ,  \qquad g_3=\sin^2\theta_1\, \sin^2\theta_2\ , \dots \ .
\ee
We will use the notation
\be
(\p_\Omega F)^2\equiv\sum_{i=1}^{d-2} {1\ov g_i}\, \le(\p F\ov \p\theta_i\ri)^2 \ ,
\ee
and denote the set of $\theta_i$'s as $\Omega$.

We parametrize the entangling surface in polar coordinates as
\be
\rho = R \, r (\Omega)
\ee
where $r (\Omega)$ specifies the shape of the surface, while $R$ gives its size. 
The minimal surface $\rho(z,\Omega)$ then satisfies the boundary condition $\rho(z=0,\Omega)=R \, r (\Omega) $. 

The entanglement entropy is given by the minimal surface area:
\be 
S (R)= {2 \pi  L^{d-1} \ov \ka^2} A =  K' A , \quad K' \equiv {2 \pi  L^{d-1} \ov \ka^2} \ ,
\ee
where
\be
A  =  \int_0^{z_t} d z \int d\Omega_{d-2} \ {\rho^{d-2} \ov z^{d-1}} \sqrt{(\p_z\rho)^2 + {1 \ov f (z)}\le(1+{(\p_\Omega \rho)^2\ov \rho^2}\ri)}   =  \int_0^{z_t} d z \int d\Omega\  \sL \ .
\ee

One can go through the same steps as for the sphere case, where $r(\Omega)=1$, to obtain the near boundary expansion:
\be  
\rho(z,\Omega) = R  \, r(\Omega)  - {z^2 \ov 2R}   \, \tilde{r}(\Omega) + \cdots + c_d (R,\Omega) z^d + \cdots + \sum_{n=2,m=2}^\infty a_{nm} (R,\Omega) z^{n+m \al} \ .
\ee
$ \tilde{r}(\Omega)$ and the functions appearing in higher orders can be determined by solving algebraic equations only involving $ r(\Omega)$ and its derivatives. One can use the asymptotic data, $c_d (R,\Omega)$ to obtain $dA/dR$, by using the Hamilton-Jacobi formalism~\cite{Liu:2012eea}. We take $z$ to be time, and introduce the canonical momentum and Hamiltonian
\bln
\Pi &= \int d\Omega\, {\p \sL \ov \p (\p_z\rho)} =  \int d\Omega\, {\rho^{d-2} \ov z^{d-1}}\,  {\rho' \ov   \sqrt{(\p_z\rho)^2 + {1 \ov f (z)}\le(1+{(\p_\Omega \rho)^2\ov \rho^2}\ri)}} \\
\sH &= \Pi \rho' - \sL = - \int d\Omega\,{\rho^{d-2} \ov z^{d-1}} {1+{(\p_\Omega \rho)^2\ov \rho^2} \ov f \sqrt{(\p_z\rho)^2 + {1 \ov f (z)}\le(1+{(\p_\Omega \rho)^2\ov \rho^2}\ri)} }
\end{align}
One can show that
\be \label{univshape}
{dA \ov dR} =  - d R^{d-2}\, c_d (R) - {\tilde{e}_d \ov R} +\dots \ ,
 \ee
 where $\tilde{e}_d$ is proportional to $e_d$ in~\eqref{eopr}, dots denote non-universal terms that drop out when acted on with the differential operator~\eqref{Scen}, and
 \be
 c_d (R)\equiv  \int d\Omega \, {r(\Omega)^{d-1} \ov \sqrt{1+{\le(\p_\theta r(\theta_i)\ri)^2\ov r(\Omega)^2}}} \, c_d (R,\Omega)\ .
 \ee 
As a result $\sS_d (R)$ can be solely expressed in terms of $c_d(R)$, and the same formulae apply as in section~\ref{sec:eesp}.

In the large $R$ limit we consider the expansion 
\be \label{Rexp1}
\rho (z,\theta_i) = R\, r(\Omega) - \rho_0 (z,\Omega) + \cdots  \ .
\ee
Plugging in the above expression into the equation of motions we can readily solve $\rho_0$ 
\be
\rho_0 (z,\Omega)= \int_0^z dz {z^{d-1} \ov f^\ha \sqrt{a(\Omega)^{-1} - z^{2 (d-1)}}}\, \sqrt{1+{\le(\p_i r(\Omega)\ri)^2\ov r^2(\Omega}}
\ee
where $a(\Omega)$ is an integration ``constant'' to be determined. As in~\eqref{onne}, considering that the UV expansion~\eqref{Rexp1} should break down precisely at the horizon, we require
 that 
 \be
 a(\Omega)=1 \ .
 \ee
 Then $\rho_0$ factorizes and we obtain:
\be
\rho_0 (z,\Omega)= \sqrt{1+{\le(\p_\Omega r(\Omega)\ri)^2\ov r^2(\Omega)}} \, \rho_0^{\le(S\ri)}= \sqrt{1+{\le(\p_\Omega r(\Omega)\ri)^2\ov r^2 (\Omega)}} \,  \int_0^z dy \, {y^{d-1} \ov f^\ha \sqrt{1 - y^{2 (d-1)}}} \ ,
\ee
where $\rho_0^{\le(S\ri)}$ is the sphere result given in \eqref{sol1}. We readily obtain:
\bln
 c_d (R,\Omega)&=-{1\ov d}\,  \sqrt{1+{\le(\p_\Omega r(\Omega)\ri)^2\ov r(\Omega)^2}}\\
c_d(R)&= \int d\Omega \, {r(\Omega)^{d-1} \ov \sqrt{1+{\le(\p_\Omega r(\Omega)\ri)^2\ov r(\Omega)^2}}} \, c_d (R,\Omega)= - {1\ov d}\int d\Omega\,  r(\Omega)^{d-1}=-{(d-1)\, V_{\Sig}\ov  d\, R^{d-1}} \ ,
\end{align}
where $V_{\Sig}$ is the volume enclosed by $\Sig$. Plugging into \eqref{imexp} yields the result
\be
S^{(\Sig)}=K'\, V_{\Sig}+\cdots \ .
\ee

\vspace{0.2in}   \centerline{\bf{Acknowledgements}} \vspace{0.2in} 
We thank S.~Pufu for conversations.  Work supported in part by funds provided by the U.S. Department of Energy
(D.O.E.) under cooperative research agreement DE-FG0205ER41360, and by a Simons Fellowship. HL also thanks the Isaac Newton Institute for Mathematical Sciences for hospitality
during the last stage of this work.

\appendix

\section{The $n=2$ case} \label{app:n2}

In the $n=2$ case the minimal surface ending on the boundary theory sphere has disk topology. This was seen before in~\cite{Liu:2012eea}, where the Coulomb branch flow of $d=4$ MSYM~\cite{Freedman:1999gk} was analyzed. 

Firstly, we analyze the IR region. In section~\ref{sec:scalingIR} we saw that the small $u$ expansion~\eqref{jepexp} of the reference solution $\bar{\rho}_d$ was singular for $n=2$. Unlike in the $n<2$ case, the expansion does not start with a constant term:
\be
\bar{\rho}_d\le(u\ri)=\sqrt{2(d-2)\ov d-1}\, \sqrt{\log {u_0\ov u}}+\dots+\bar{h\ov u_0^{d-1}}\, u^{d-1}+\dots \qquad (u\to 0) \ .
\ee
 \eqref{jep} then implies that $\rho(z)$ has the small $z/z_t$ expansion valid in the region~\eqref{ovp1}:
\be\label{IRn2}
\begin{split}
\rho(z)&=\sqrt{2(d-2)\ov (d-1)a}\, \sqrt{\log { u_0\,  z_t\ov z}}+\dots+\bar{h\ov \sqrt{a}\, (u_0\,  z_t)^{d-1}}\, z^{d-1}+\dots\\
&=\sqrt{2(d-2)\ov (d-1)a}\, \sqrt{\log ( \sqrt{a}\, u_0\,  z_t)}\le[1-\ha\, {\log \le(\sqrt{a}\,z\ri)\ov \log ( \sqrt{a}\, u_0\,  z_t)}+\dots\ri] +\bar{h\ov \sqrt{a}\,( u_0\,  z_t)^{d-1}}\, z^{d-1}+\dots \ .
\end{split}
\ee

Let us turn our attention to the UV expansion~\eqref{exp}. We have to modify it so that $\hat{\rho}$ is multiplied by a general function $F(R)$, not $R^{-\nu}$. To obtain the large $z$ behavior of $\rho_1(z)$ we go through the same steps as in~\eqref{rhi1} to get:
\be
\rho_1(z)={b_1\ov (d-1)\sqrt{a}}\,z^{d-1}\le(1+\dots\ri)+{d-2\ov (d-1) a}\, \log z\le(1+\dots\ri) \ .
\ee
We note that taking the $n\to2$ limit of~\eqref{rhi1} can also give us this result. Plugging in $n=2$ into~\eqref{gje1}, and combining all this together in~\eqref{exp} gives:
\be
\rho(z)=R-{1\ov (d-1)\sqrt{a}}\,\le({b_1\ov R}+\dots+{\hat{b}}\,F(R)\ri) \,z^{d-1}-{d-2\ov (d-1) a \, R}\, \log z+\dots \ .
\ee

Matching this expansion to the IR solution~\eqref{IRn2} determines
\bea
R&=&\sqrt{2(d-2)\ov (d-1)a}\, \sqrt{\log(\sqrt{a}\,  u_0\,  z_t)} \\
b&=&0\\
\hat{b}&=&-(d-1)\, \bar{h}\\
F(R)&=&a^{(d-1)/2}\,\exp\le(-{(d-1)^2 \, a\ov 2(d-2)}\, R^2\ri) \ .
\eea
It would be very interesting where exponential behavior comes from in field theory.

The non-analytic contribution to $\sS_d$ is also exponentially small for $n=2$. Using~\eqref{eopr} and~\eqref{uwme} the leading large $R$ contribution we get for 
$\sS_d$ is
\be
\sS_d^{\text{(non-analytic)}}\propto \le(a\,R^2\ri)^{t}\, \exp\le(-{(d-1)^2 \, a\ov 2(d-2)}\, R^2\ri) \qquad t\equiv{d-3\ov 2}+\le[d\ov 2\ri] \ . \label{n2nonanal}
\ee

\section{$1/R$ term in the $d=3$ scaling geometries}\label{app:scalingonshell}

Let us divide the area func   an IR part and let $z_*$ be some $z$ in the matching region that divides between the two regions. It is clear that the result should not depend on $z_*$.
\be\label{Adivision}
A\equiv A_{UV}+A_{IR}=\int_0^{z_*} dz \ {\rho \ov z^{2}} \sqrt{\rho'(z)^2+{1\ov f(z)}}+\int_{z_*}^{z_t} dz \ {\rho \ov z^{2}}   \sqrt{\rho'(z)^2+{1\ov f(z)}}\ .
\ee
For $A_{UV}$, we can go through the same steps leading to~\eqref{AOnShell2}. We obtain 
\be
A_{UV}=\#\, R+{1\ov R} \int_0^{z_*} dz \ \le[{\sqrt{f(z)}\ov 2 z^2} \rho_1'(z)^2 - {\rho_1(z)\ov  z^2} \ri]+O\le({1\ov R^3}\ri)  \ .
\ee
Because the integrand for $A_{UV}$ is the same as in the first line of~\eqref{AOnShell1}, and only the upper limit of the integral differs, in analogy with~\eqref{AOnShell1} and~\eqref{AOnShell2}, we obtain
\be\label{AUVscaling}
A_{UV}=\# \, R+{1\ov R}\le[-\ha \int_0^{z_{*}} dz \ {z^2 \ov \sqrt{f(z)}} \le[\int_z^\infty dv \ {1\ov v^2 \sqrt{f(v)}}\ri]^2+\ {\sqrt{f(z)}\ov z^2} \rho_1'(z)\rho_1(z)\Big\vert_{z=z_{*}}\ri]+\dots \ ,
\ee
where the last term is a boundary term that vanished in~\eqref{AOnShell2}; here it will play an important role.

For $z>z_{CO}$ we will assume for simplicity that $f(z)=a\, z^n$ exactly. We set $a=1$ to avoid clutter. Corrections to $f(z)$ can be understood in a perturbative setup, and for fast enough convergence to the asymptotic behavior, the results obtained below should hold. In Appendix~\ref{dwonshell}, we show how to incorporate subleading terms in $f(z)$ for domain wall flows. 
Because we have the full scaling symmetry in the IR, we can evaluate the IR on-shell action by using the solution $\bar{\rho}_d(z)$ introduced in~\eqref{jep}.
\be
A_{IR}=\int_{z_*}^{z_t} dz \ {\rho \ov z^{2}}   \sqrt{\rho'(z)^2+{1\ov f(z)}}={1\ov  z_t^{n}}\, \int_{z_*/z_t}^{1} du \, {\bar{\rho}_d(u) \ov u^{2}}   \sqrt{\bar{\rho}'_d(u)^2+{1\ov u^n}} \ .
\ee
For small $u$ we can plug in the UV expansion~\eqref{jepexp} of $\bar{\rho}_d(z)$ into the integral to obtain the leading behavior of the integrand
\bea
{\bar{\rho}_d(u) \ov u^{2}}   \sqrt{\bar{\rho}'_d(u)^2+{1\ov u^n}}&=&{\bar{\al}_0\ov u^{2+n/2}}+{\al_1+{(2-n)^2\ov 2} \al_1^2\ov \bar{\al}_0 \, u^{3n/2}}+\dots\cr
&&+{\bar{h} \le(1+\le(\frac2n - 1\ri) \le(\frac6n - 1\ri) \al_1\ri)\ov \bar{\al}_0^{2 \ov \eta}} \, u^{1-n/2}+\dots \ . \label{divergences}
\eea
We have to subtract the divergences from the integrand coming from the first line of~\eqref{divergences}, in order to be able to obtain the $1/R$ expansion of $A_{IR}$. Note that for $n<2/3$, only the first term gives a divergence. For $2/3<n<4/5$, only the first two terms give a divergence, and so on. It does not hurt to subtract arbitrary regular terms from the integrand, so we can proceed by subtracting the first few terms in the first line of~\eqref{divergences}. Finally, we can write down the result for $A_{IR}$

\bea
A_{IR}&=&{1\ov z_t^{n}}\, \int_{z_*/z_t}^{1} du \, \le[ {\bar{\rho}_d(u) \ov u^{2}}   \sqrt{\bar{\rho}'_d(u)^2+{1\ov u^n}}-{\bar{\al}_0\ov u^{2+n/2}}-{\al_1+{(2-n)^2\ov 2} \al_1^2\ov \bar{\al}_0 \, u^{3n/2}}+\dots\ri]\cr
&&+{1\ov z_t^{n}}\, \le[-{\bar{\al}_0\ov \le(1+\frac n2\ri)\, u^{1+n/2}}-{\al_1+{(2-n)^2\ov 2} \al_1^2\ov \le({3n\ov2}-1\ri)\,\bar{\al}_0 \, u^{3n/2-1}}+\dots\ri]_{u=z_*/z_t}^{u=1} \ .\label{AIRresult}
\eea
For $n=2/3$ the above equation is replaced by
\bea
A_{IR}&=&{1\ov z_t^{2/3}}\, \int_{z_*/z_t}^{1} du \, \le[ {\bar{\rho}_d(u) \ov u^{2}}   \sqrt{\bar{\rho}'_d(u)^2+{1\ov u^{2/3}}}-{\bar{\al}_0\ov u^{7/3}}-{\al_1+\frac89 \al_1^2\ov \bar{\al}_0 \, u}\ri]\cr
&&+{1\ov z_t^{2/3}}\, \le[-{\bar{\al}_0\ov \frac43\, u^{4/3}}+{\al_1+\frac89 \al_1^2\ov \bar{\al}_0}\,\log\, u\ri]_{u=z_*/z_t}^{u=1} \ .
\eea
The lower limit of the integral in the first line can be sent to zero without encountering divergences. Using~\eqref{893} we can trade $z_t$ for $R$
\be
z_t=\le({R\ov\bar{\al}_0}\ri)^{2/(2-n)} \ .
\ee
 We obtain
\bea
A_{IR}={\#\ov  R^{2n/(2-n)}}+{ R\ov  \le(1+\frac n2\ri)\, z_*^{1+n/2}}+ {1\ov R}\, {\al_1+{(2-n)^2\ov 2} \al_1^2\ov \le({3n\ov2}-1\ri)\,z_*^{3n/2-1}}+\dots \ ,
\eea
where the expansion is a double expansion as in~\eqref{ovp1}. For $n=2/3$ the answer is:
\be
A_{IR}={\#\ov  R}+{ R\ov  \frac43\, z_*^{1+n/2}}-{27\ov 64}\, {\log R\ov R}+\dots \label{n23result}
\ee

 We know the coefficient of the first term from the analysis performed in the main text. In this approach it is given by a more complicated expression: the integral in the first line (with the lower limit sent to zero) and the $u=1$ boundary terms in the second line in~\eqref{AIRresult}. It is related to $e_n$ by some simple factors. The second term is an uninteresting area law term. The third term is the $1/R$ term we are after. Combining this term with the boundary term in~\eqref{AUVscaling} we get for the $1/R$ term:
\be
A=\#\, R+{\#\ov R^{2n/(2-n)}}-{1\ov R}\le[\ha \int_0^{z_{*}} dz \ {z^2 \ov \sqrt{f(z)}} \le[\int_z^\infty dv \ {1\ov v^2 \sqrt{f(v)}}\ri]^2+{2 \ov (3n/2-1) (2 + n)^2}\, { 1\ov z_*^{3n/2-1}}\ri]+\dots \ , 
\ee
where we plugged in the value of $\al_1$~\eqref{a1} and the UV expansion of $\rho_1$~\eqref{rhi1}. For $n>2/3$ the two terms beautifully combine to give:
\bea
A&=&\#\, R+{\#\ov R^{2n/(2-n)}}-{a_1\ov 2R}+\dots\\
a_1&=&\int_0^{\infty} dz \ {z^2 \ov \sqrt{f(z)}} \le[\int_z^\infty dv \ {1\ov v^2 \sqrt{f(v)}}\ri]^2 \ . 
\eea
For $n=2/3$ there are no terms coming from~\eqref{AUVscaling} that could contribute to the $\log R/R$ term of~\eqref{n23result}. Hence we obtain:
\be
A=\#\, R-{27\ov 64}\, {\log R\ov R}+{\#\ov  R}+\dots \ . \label{n23result2}
\ee 
For $n<2/3$ we have to apply subtractions, then $a_1$ is given by
\be
a_1=\int_0^{\infty} dz \  \le({z^2 \ov \sqrt{f(z)}}\le[\int_z^\infty dv \ {1\ov v^2 \sqrt{f(v)}}\ri]^2-{4 \ov (2 + n)^2}\, { 1\ov z^{3n/2}}\ri) \ . 
\ee
Note that in the main text we use a dimensionless version of $a_1$ denoted by $s_1$. Because we set $a=1$ in this appendix, plugging in $\tilde\mu=1$ in the expression of $s_1$ gives the result for $a_1$ obtained here.

\section{Details of the UV expansion of $\rho_1$ for the domain wall case} \label{sec:exp}

We are interested in the behavior of $\rho_1$ at large $z$ beyond the crossover scale $z_{CO}$:  $z_{CO}\ll z\ll R$. We assume that $f(z)$ takes the form: 
\be\label{faltform}
f(z)=f_\infty\le(1-{\lam \ov z^{2\tilde\alpha} }\ri)+\dots \qquad (z \gg z_{CO}) \ .
\ee
where we introduced $\lam\equiv\tilde\mu^{-2\tilde\al}$.
From~\eqref{rho1s}
\be 
\rho_1 = b_1 \rho_{hom} (z)+ \xi (z)
\ee
with 
\be 
\xi (z) \equiv(d-2)\int_{0}^z du \ {u^{d-1}\ov \sqrt{f(u)}} \int_{u}^\infty dv \ {1 \ov v^{d-1}\sqrt{f(v)}} \ .
\ee
For large $z$, $\rho_{hom} (z) $ has the expansion
\be 
\rho_{hom} (z) = {z^d \ov d \sqrt{f_\infty}} \le(1+ { d \lam \ov 2 (d-2\tilde \al)} z^{-2\tilde\al}+\dots\ri) + O(z_{CO}^d) \ .
\ee
The large $z$ behavior of $\xi (z)$ is a bit more complicated. For $\tilde \al > 1$ we have 
\be
\begin{split}
\xi(z) =& \int_{0}^z du \ \le[{u^{d-1}\ov \sqrt{f(u)}} (d-2)\int_{u}^\infty dv \ {1 \ov v^{d-1}\sqrt{f(v)}}-{u\ov f_\infty}\ri]+{z^2\ov 2 f_\infty}\\
=&{z^2\ov 2 f_\infty} + \ga - \int_{u}^\infty du \ \le[{u^{d-1}\ov \sqrt{f(u)}} (d-2)\int_{u}^\infty dv \ {1 \ov v^{d-1}\sqrt{f(v)}}-{u\ov f_\infty}\ri] \cr
= & {z^2\ov 2 f_\infty} + \ga -  {\lam \ov 2 f_\infty}{(d-2+\tilde\al) \ov (\tilde\al-1)(d-2+2\tilde\al)} z^{2-2\tilde\al} + O( z^{2-4\tilde\al})
\end{split}
\ee
with 
\be 
\ga =  \int_{0}^\infty du \ \le[{u^{d-1}\ov \sqrt{f(u)}} (d-2)\int_{u}^\infty dv \ {1 \ov v^{d-1}\sqrt{f(v)}}-{u\ov f_\infty}\ri], \qquad \tilde \al > 1 \ .
\ee
For $1 \geq \tilde\al>1/2$ we have to do more subtractions:
\be
\begin{split}
\xi(z)=&\int_{0}^z du \ \le[{u^{d-1}\ov \sqrt{f(u)}} (d-2)\int_{u}^\infty dv \ {1 \ov v^{d-1}\sqrt{f(v)}}-{u\ov f_\infty}-{ \lam (d-2+\tilde\al) \ov(d-2+2\tilde\al)f_\infty} u^{1-2\tilde\al}\ri] \\
& +{z^2\ov 2 f_\infty}\le(1+ { \lam (d-2+\tilde\al) \ov (1-\tilde\al)(d-2+2\tilde\al)} z^{-2\tilde\al}\ri)\\
=& {z^2\ov 2 f_\infty}\le(1+ { \lam (d-2+\tilde\al) \ov (1-\tilde\al)(d-2+2\tilde\al)} z^{-2\tilde\al}\ri)+ \ga + O\le(z^{2-4\tilde\al}\ri) \ ,
\end{split}
\label{oeep}
\ee 
where now $\ga$ is given by
\be
\ga = \int_{0}^\infty du \ \le[{u^{d-1}\ov \sqrt{f(u)}} (d-2)\int_{u}^\infty dv \ {1 \ov v^{d-1}\sqrt{f(v)}}-{u\ov f_\infty}-{ \lam (d-2+\tilde\al) \ov(d-2+2\tilde\al)f_\infty} u^{1-2\tilde\al}\ri]  \ .
\ee

For $\tilde \al$ outside the above ranges one has to do more subtractions, but the leading expressions remain the same as~\eqref{oeep} with the explicit value of $\ga$ being different.


\section{$1/R$ term in the $d=3$ domain wall geometry}\label{dwonshell}


In the domain wall case we follow the same logic as in Appendix~\ref{app:scalingonshell}, i.e.~we divide the area functional into UV and IR parts as in~\eqref{Adivision}. The UV expansion for scaling and domain wall geometries takes the same form, and correspondingly $A_{UV}$ has an identical form to~\eqref{AUVscaling}. $z_*$ is an arbitrary point in the region~\eqref{mheo}.

$A_{IR}$ can be obtained by regarding $f(z)$ as a perturbation of $f_\infty$  and working to first order. We set up the IR problem a bit differently, than in section~\ref{sec:domain}:
\bea
\rho(z)&=&r_0(z)+\lam r_1(z)=\sqrt{R^2-{z^2\ov f_\infty}}+\lam r_1(z)\\
\lam r_1(z)-{z^2\ov 2f_\infty R}&=&-{\rho_1(z)\ov R}+\dots \ , \label{r1rho1}
\eea
where $\lam=\tilde\mu^{-2\tilde\al}$ as in~\eqref{faltform}, and the above equation follows from 
\be
R-{\rho_1(R)\ov R}+\dots=\sqrt{R^2-{z^2\ov f_\infty}}+\lam r_1(z)+\dots \ .
\ee

Let us consider how the on-shell action $A_{IR}$ changes, if we change $f(z)$.  If we regard $z$ as time, this is as a Hamilton--Jacobi problem in classical mechanics, when we are interested in how the on-shell action changes. In this analogy, we are holding the initial time and the endpoint of the trajectory fixed. There will be a term coming from the explicit change of $f(z)$ in the Lagrangian. Because the original trajectory was an extremum of the action there is only a boundary term coming from the change of trajectory. Finally, there is a term coming from the change of time, when the particle reaches the endpoint. Hence we get, in the order we listed the terms above:
\be 
\de A_{IR} = \int_{z_*}^{z_{t}} d z \,  {\de \sL \ov \de f} \de f  
 - \Pi \,\de \rho  \Big\vert_{z_*} - \sH (z_t) \de z_t \ ,
\ee
where $\de z_m$ and $\de \rho$ denote the induced variations due to $\de f$, and the canonical variables have the expressions
\be \label{canoD}
\Pi = {\p \sL \ov \p \rho'} =  {\rho^{d-2} \ov z^{d-1}} {\rho' \ov \sqrt{\rho'^2 + {1 \ov f}}}, \qquad
\sH = \Pi \rho' - \sL = - {\rho^{d-2} \ov z^{d-1}} {1 \ov f \sqrt{\rho'^2 + {1 \ov f}}} \ .
\ee
Applying the above results to the current problem, we find that 
\be \label{w0}
\de A_{IR}  = \int_{z^*}^{z_m} dz \,  {\de \sL \ov \de f}\biggr|_{r_0}  \, \le(-{f_\infty \lam\ov z^{2\tilde\al}}\ri)  -\Pi (z_*) \Big\vert_{r_0} \lam r_1 (z_*)
\ee
where we used $\sH (z_t)=0$. Evaluating these with $z_m=\sqrt{f_\infty}\, R$ we get:
\be
\de A_{IR}  =\#\, R- f_\infty^{-(1+\tilde\al)} \lam \, {R^{-2\tilde\al}\ov 1-4\tilde\al^2}+{\lam\ov 2(1-2\tilde\al)f_\infty^{3/2}\,R}\, z_*^{1-2\tilde\al}+{1\ov \sqrt{f_\infty}\, z_*} \lam r_1 (z_*)+\dots \ .
\ee
The zeroth order contribution gives:
\be
A_{IR}^{(0)}={R\ov \sqrt{f_\infty} z_*}-{1\ov f_\infty}=\#\, R-{1\ov f_\infty} \ .
\ee
Adding all this up and using \eqref{r1rho1} we get: 
\be
\begin{split}
A=&\#\, R-{1\ov f_\infty} - f_\infty^{-1-\tilde\al} \lam \, {R^{-2\tilde\al}\ov 1-4\tilde\al^2}- {1\ov 2R} \int_0^{z_{*}} dz \ {z^2 \ov \sqrt{f(z)}} \le[\int_z^\infty dv \ {1\ov v^2 \sqrt{f(v)}}\ri]^2\\
&+\ {\sqrt{f(z_*)}\ov z_*^2 R} \rho_1'(z_*)\rho_1(z_*)
+{\lam\ov 2(1-2\tilde\al)\sqrt{f_\infty}\,R}\, z_*^{1-2\tilde\al}-{1\ov\sqrt{f_\infty}\, z_*\, R} \rho_1 (z_*)+{z_*\ov 2 f_\infty^{3/2}\, R} \ . \label{almost}
\end{split}
\ee
Note that this result is in the double expansion~\eqref{mheo}, just like all expressions in the matching region appearing in the main text. Now the common theme of this paper has to be applied: subtractions. Subtracting the divergence(s) from the integral allows us to go with the upper limit to infinity and gives the result:
\bea
A&=&\#\, R-{1\ov f_\infty} - f_\infty^{-1-\tilde\al} \lam \, {R^{-2\tilde\al}\ov 1-4\tilde\al^2}- {a_{1}\ov 2R}  \label{almost2}\\
a_{1}&=&\begin{cases}
\int_0^{\infty} dz \ \le[{z^2 \ov \sqrt{f(z)}} \le[\int_z^\infty dv \ {1\ov v^2 \sqrt{f(v)}}\ri]^2-{1 \ov f_\infty^{3/2}} \ri] \qquad\qquad\qquad \le(\ha<\tilde\al\ri) \\
\int_0^{\infty} dz \ \le[{z^2 \ov \sqrt{f(z)}} \le[\int_z^\infty dv \ {1\ov v^2 \sqrt{f(v)}}\ri]^2-{1 \ov f_\infty^{3/2}}\le(1+{3+2\tilde\al\ov 2(1+2\tilde\al)}\, {\lam\ov z^{2\tilde\al}}\ri) \ri] \qquad\qquad \le(\frac14<\tilde\al<\ha\ri)
\end{cases} \label{dwresult}
\eea
The rest of the terms in \eqref{almost} (after adding back the subtracted part to the integral) can be shown to cancel to the order in $z_*/R$ and $1/(\tilde\mu z_*)$ that we wrote them down.

The final answer is:
\be
\sS_3=\sS_3^{(IR)}+K_{IR} \, { f_\infty^{-\tilde\al} \lam\ov (1-2\tilde\al)R^{2\tilde\al}}+{K s_{1}\ov \tilde\mu R}+\dots \ ,
\ee
where $s_1=\tilde\mu a_1$, as explained around~\eqref{a1e}.

\section{Some results for closely separated fixed points} \label{sec:close} 

We review and extend some results from \cite{Liu:2012eea} for closely separated fixed points:
\bea
f(z)&=&1+\ep g(z) \qquad g(z)\to 1- {\lam\ov z^{2\tilde\al}} \quad (z\to\infty)  \label{fepexp}\\
\sS_d&=&\sS_d^{UV}-\ep\, {(d-1)!! K\ov 2(d-2)!!}\begin{cases}
  \int_0^1 dx \ g(x R) \qquad \text{$d$ odd}\\
    \int_0^1 dx \ {x g(x R)\ov \sqrt{1-x^2}} \qquad \text{$d$ even}.
\end{cases}
\eea
Let us start with the odd $d$ case and expand for large $R$ with the technique of subtraction:
\be
\begin{split}
\int_0^1 dx \ g(x R)&=1+\int_0^1 dx \ \le[g(x R)-1\ri]=1+\le[xg(xR)\ri]_{x=0}^1+\int_0^1 dx \ \le[-x g'(x R) R-1\ri]\\
&=1-\int_0^1 dx \ (x R) g'(x R)+\dots\\
&=1-\le[\int_0^\infty  dz\  z g'(z) \ri]\,{1\ov R}+\dots \ ,
\end{split}
\ee
where we used partial integration and assumed fast enough ($\tilde\al>\ha$) decay at infinity. If the decay is slower, we need additional subtractions. For the even dimensional case we encounter an integral similar to \eqref{rzt2}, so we can use the approximation technique from there. After subtraction the integral is expected to be dominated by the $x\ll 1$ region and we have:
\be
\begin{split}
\int_0^1 dx \ {x g(x R)\ov \sqrt{1-x^2}}&=1+\int_0^1 dx \ {x g(x R)-x\ov \sqrt{1-x^2}}=1+\int_0^1 dx \ \le[x g(x R)-x\ri]+\dots\\
&=1-\ha \int_0^1 dx \ x^2 g'(x R) R+\dots\\
&=1-\ha\le[\int_0^\infty  dz\  z^2 g'(z) \ri]\,{1\ov R^2}+\dots \ .
\end{split}
\ee

The final result in odd $d$ is:
\bea
\sS_d&=&\sS_d^{UV}-\ep\, {(d-1)!! K\ov 2(d-2)!!}+\ep\,K\,\ha \,{(d-1)!!\ov (d-2)!!}\,  b(\tilde\al) \, {\lam\ov R^{2\tilde\al}}+\ep\, {K s_1\ov \tilde\mu R}+\dots\\
s_{1}&=&{(d-1)!!\, \tilde\mu \ov 2(d-2)!!}\,\begin{cases}
\int_0^\infty  dz\  z g'(z) \qquad\qquad\qquad\quad (\tilde\al>\ha)\\
\int_0^\infty dz\  \le[z g'(z)-{2\tilde\al \lam \ov  z^{2\tilde\al} } \ri] \qquad (\tilde\al<\ha) \ ,
\end{cases} \label{closely}
\eea
where $\lam=\tilde\mu^{-2\tilde\al}$.
Of course we might need to apply more subtractions, if $\tilde\al$ is small enough.

The final result for even $d$ takes the form:
\bea
\sS_d&=&\sS_d^{UV}-\ep\, {(d-1)!! K\ov 2(d-2)!!}+\ep\,K\,\ha \,{(d-1)!!\ov (d-2)!!}\,  b(\tilde\al) \, {\lam\ov R^{2\tilde\al}}+\ep\, {K s_{2}\ov (\tilde\mu R)^2}+\dots\\
s_{2}&=&{(d-1)!! \,\tilde\mu^2\ov 4(d-2)!!}\,\begin{cases}
\int_0^\infty  dz\  z^2 g'(z)  \qquad\qquad\qquad\quad (\tilde\al>1)\\
\int_0^\infty  dz\  \le[z^2 g'(z) -{2\tilde\al\lam \ov z^{2\tilde\al-1}}  \ri] \qquad (\tilde\al<1) \ .
\end{cases} 
\eea

Let us compare~\eqref{closely} to~\eqref{dwresultsmain}. We are interested in $s_1$ to  first order in $\ep$, which we repeat here for convenience  for $\ha<\tilde\al$:
\be
s_1=\int_0^{\infty} dz \ \le[{z^2 \ov \sqrt{f(z/\tilde\mu)}} \le[\int_z^\infty dv \ {1\ov v^2 \sqrt{f(v/\tilde\mu)}}\ri]^2-{1 \ov f_\infty^{3/2}} \ri]  \ .
\ee
Let us first take the integral over $v$. Using~\eqref{fepexp} we obtain:
\be
\int_z^\infty dv \ {1\ov v^2 \sqrt{f(v/\tilde\mu)}}={1\ov z}-{\ep\ov 2}\, \int_z^\infty dv \ {g(v/\tilde\mu)\ov v^2} \ .
\ee
The next step is to examine the full integrand:
\pagebreak
\bea
{z^2 \ov \sqrt{f(z/\tilde\mu)}} \le[\int_z^\infty dv \ {1\ov v^2 \sqrt{f(v/\tilde\mu)}}\ri]^2&=&z^2\le(1-{\ep\ov 2}\,g(z/\tilde\mu)\ri)\le({1\ov z^2}-{\ep\ov z}\, \int_z^\infty dv \ {g(v/\tilde\mu)\ov v^2}\ri)\cr
&=&1+\ep\,\le(-\ha\, g(z/\tilde\mu)+z\, \int_z^\infty dv \ {g(v/\tilde\mu)\ov v^2}\ri)\\
{1 \ov f_\infty^{3/2}}&=&1-{3\ep\ov 2} \ .
\eea
Combining the above terms we get that $s_1$ has the expression to first order in $\ep$:
\be\label{intermediateform}
s_{1}=\ep\, \int_0^{\infty} dz \ \le[-\ha\, g(z/\tilde\mu)+z\, \int_z^\infty dv \ {g(v/\tilde\mu)\ov v^2}+{3\ov 2} \ri]  \ .
\ee
We can define a new function $\tilde{g}(z)\equiv g(z)-1$ that vanishes sufficiently fast as $z\to\infty$. In terms of this new function
\be
\begin{split}
s_{1}&=\ep\, \int_0^{\infty} dz \ \le[-\ha\, \tilde{g}(z/\tilde\mu)+z\, \int_z^\infty dv \ {\tilde{g}(v/\tilde\mu)\ov v^2} \ri] \\
&=- \ep\, \int_0^{\infty} dz \ \tilde{g}(z/\tilde\mu)\\
 &=\ep\,\tilde\mu \int_0^{\infty} dz \ zg'(z)\ ,
\end{split}
\ee
where in the second line we integrated the second term partially in $z$. In the third line we did a second partial integration in $z$, and used that $\tilde{g}'(z)\equiv g'(z)$. 

For $\tilde\al<\ha$ the same steps lead to the subtracted version of~\eqref{intermediateform}:
\be
s_{1}=\ep\, \int_0^{\infty} dz \ \le[-\ha\, g(z/\tilde\mu)+z\, \int_z^\infty dv \ {g(v/\tilde\mu)\ov v^2}+{3\ov 2}- {3+2\tilde\al\ov 2(1+2\tilde\al)}\, {1\ov z^{2\tilde\al}} \ri]  \ .
\ee
Defining $\tilde{g}(z)\equiv g(z)-1+{\lam\ov z^{2\tilde \al}}$ allows us to absorb all the subtracted terms, and get the simple formula:
\be
s_{1}=- \ep\, \int_0^{\infty} dz \ \tilde{g}(z/\tilde\mu) \ .
\ee
Partially integrating in $z$ and using $\tilde{g}'(z)\equiv g'(z)-{2\tilde\al\, \lam\ov z^{2\tilde \al+1}}$ we obtain~\eqref{closely}.

\end{document}